\documentclass[aps,reprint,superscriptaddress]{revtex4-2}

\usepackage{amsmath,amssymb}
\usepackage{graphicx}
\usepackage{bm}
\usepackage[hidelinks]{hyperref}
\usepackage{ragged2e}

\newcommand{\pd}{\partial}

\begin{document}

\title{Chemotactic recruitment offsets congestion in honeybee trophallaxis}

\author{William Magrogan}
\affiliation{Department of Applied Mathematics, University of Colorado Boulder, Boulder, Colorado, USA}

\author{Orit Peleg}
\affiliation{BioFrontiers Institute, University of Colorado Boulder, Boulder, Colorado, USA}
\affiliation{Santa Fe Institute, Santa Fe, New Mexico, USA}

\author{Zachary P. Kilpatrick}
\email{zpkilpat@colorado.edu}
\affiliation{Department of Applied Mathematics, University of Colorado Boulder, Boulder, Colorado, USA}
\affiliation{National Institute for Theory and Mathematics in Biology, Chicago, Illinois, USA}

\date{\today}

\begin{abstract}
Honeybee colonies distribute food by trophallaxis, a sequence of local interactions in which nectar donors transfer resources to hungry nestmates. Efficient transfer requires a searcher to find a donor and then be served. Chemical signals may draw searchers to a donor, though the crowd they form may block the way. We develop a stochastic framework coupling velocity-jump search that depends on chemical signaling to random-order service at donor aggregations, and measure trophallactic efficiency by the time the colony needs to feed a prescribed fraction of its members. Directed recruitment emerges because fed bees linger near donors and strengthen local chemical cues, but lingering also slows service. Trophallactic efficiency is therefore maximized at an intermediate abandonment rate that balances recruitment against congestion. We determine when displaced cues remain useful and how competing donors partition searchers. With two donors of unequal service rate, the faster one produces fed bees sooner and so advertises more strongly, drawing recruitment toward the donor that can sustain more throughput and away from the one that would congest. Together, these results show how feedback between service and signaling can self-organize recruitment to match local service capacity without centralized coordination.
\end{abstract}

\maketitle

\section{Introduction}
\label{sec:intro}

Eusocial insects coordinate search, communication, and resource exchange without centralized control~\cite{mlot_dynamics_2012,benton_altruistic_1992,lihoreau_collective_2017,shishkov_social_2022}. Division of labor, where some individuals forage for food while others are performing tasks inside the nest, is one of the defining features of eusocial colonies. Upon return to the nest, foragers may deposit the food for long-range storage, or share it immediately with others in the nest, depending on the colony’s nutritional state, the quality of the food, and the availability of storage. In honeybee colonies, nectar is passed directly to nestmates through trophallaxis, either to receiver bees who process and store it or to food-deprived (henceforth, hungry bees) who consume it directly. We focus on the latter, where the rate at which food spreads through the colony depends on both the motion of hungry recipients and the local organization that develops around donors~\cite{gil_foraging_2002,gharooni-fard_data-driven_2020,gharooni-fard_efficiency_2020,lemanski_multiscale_2019}. 

Automated tracking now resolves these exchanges individually, in bees~\cite{gernat_automated_2018} and in ants where labeled food can be followed through the colony~\cite{greenwald_ant_2015,greenwald_individual_2018}, and shows a distribution process that is bursty, heterogeneous, and regulated by local rules rather than by any global accounting. These processes couple individual movement to a changing information environment because successful interactions may create fed bees that remain near a donor and contribute to local recruitment cues. The cue we consider is a secretion from the Nasonov gland~\cite{pickett_nasonov_1980}, which foragers expose at a food source, with the onset and duration of exposure set by the profitability of the source~\cite{free_role_1972,scenting_ref,fernandez_changes_2001,fernandez_reward_2003}. Scenting bees fan their wings to direct pheromone backward along their bodies, and in swarms locating a queen, bees that detect the signal stop and scent in turn, so the signal propagates well beyond the range of a single bee~\cite{nguyen_flow-mediated_2021,nguyen_robustness_2021,nguyen_shortest_2023}. We assume that fed bees scent while they linger at a donor, as foragers do at food sources and bees do in swarms, consistent with our experiments on group trophallaxis~\cite{gharoonifard_inprep}. We then quantify how this signaling shapes the dynamics of trophallaxis.

A hungry bee must first find a donor offering nectar within the hive. Random walks and velocity-jump processes are standard models of search for sparse targets~\cite{bhattacharya_collective_2014,bidari_hive_2021,mezey_visual_2024,random_search_review,random_walk_intro,angelani_first-passage_2014,othmer_diffusion_2002,goldstein_seminal}. Detection is rarely perfect, and is often represented by a partially absorbing boundary or probabilistic capture~\cite{erban_reactive_2007,singer_partially_2008,grebenkov_paradigm_2020}. Whether a searcher does better to move ballistically or to scan slowly depends upon the geometry of the environment and whether any signal guides it toward the target~\cite{benichou_intermittent_2011,redner_guide_2001}. For instance, if bees near a food donor scent, their pheromone provides approximate spatial information about the donor's location. In a population, if scenter positions are spread about the donor, then the plume peak estimates that location with an error that falls as the number of scenters grows. A useful theory therefore has to distinguish the intrinsic efficiency of an unbiased search from the benefit and possible cost of following such imperfect signals.

Trophallaxis is a pairwise exchange between a donor and one hungry bee, and its duration varies widely~\cite{choi_individual_2020,gernat_automated_2018}, so a bee arriving at a donor may wait among those already gathered there~\cite{seeley_social_1989,seeley_tovey_1994}. Some are still waiting to be fed; others have been fed and not yet left. Lingering fed bees that continue to contribute to the local scent plume can make the donor easier to find while increasing crowding. Related tradeoffs between accumulation and throughput arise throughout biology. Crowding that suppresses intake is consumer interference in ecology~\cite{beddington_mutual_1975,deangelis_model_1975}, queueing explains correlations between substrates competing for shared molecular processors~\cite{cookson_queueing_2011,mather_correlation_2010}, and motor traffic on a microtubule jams once dissociation from the end is slow enough to produce a pileup~\cite{leduc_molecular_2012}. Search-and-capture models, meanwhile, typically supply their targets with searchers from an external stream or source~\cite{bressloff_kim_2019,bressloff_competition_2020,bressloff_queueing_2020,giral_gmc_2025}. Here, by contrast, the population accumulating near a donor can itself contribute to the signal that draws subsequent arrivals.

Congestion at a shared resource is often self-limiting because increasing load suppresses further use. Nectar foragers, for example, gauge colony demand from how long they wait for a receiver, so queueing delay regulates subsequent foraging effort~\cite{seeley_social_1989}. Ant colonies manage severe congestion in similar ways. On a trail, flow continues to rise with density well beyond the point at which human crowds or car traffic would stall~\cite{poissonnier_experimental_2019}, whereas ants confined to a narrow space can jam despite congestion-managing behaviors such as exchanging access through social interactions~\cite{gravish_glass-like_2015}. During tunnel excavation, throughput is highest when most workers remain outside rather than crowding the excavation face~\cite{aguilar_collective_2018}. Across these examples, occupancy of a contested resource limits throughput and can generate negative feedback on further participation, as in queueing systems more generally~\cite{kleinrock_queueing_1975}.

The recruitment cue considered here creates the opposite feedback. Because the cue is produced by bees accumulating near a donor, greater occupancy can attract still more searchers, amplifying rather than damping the local load while the cue persists. Yet a plume can guide search only if it persists long enough for searchers to respond to it. Thus, bees must linger long enough to sustain a coherent recruitment cue, but leave quickly enough to relieve congestion at the donor. How long should a fed bee stay, and does the cue she helps sustain ever repay the crowding it adds? To address these questions, we develop a minimal spatial model of trophallaxis in which congestion and signaling interact, measuring efficiency as the time to feed a fraction of the colony. 

\begin{figure*}[t!]
\centering
\includegraphics[width=0.85\textwidth]{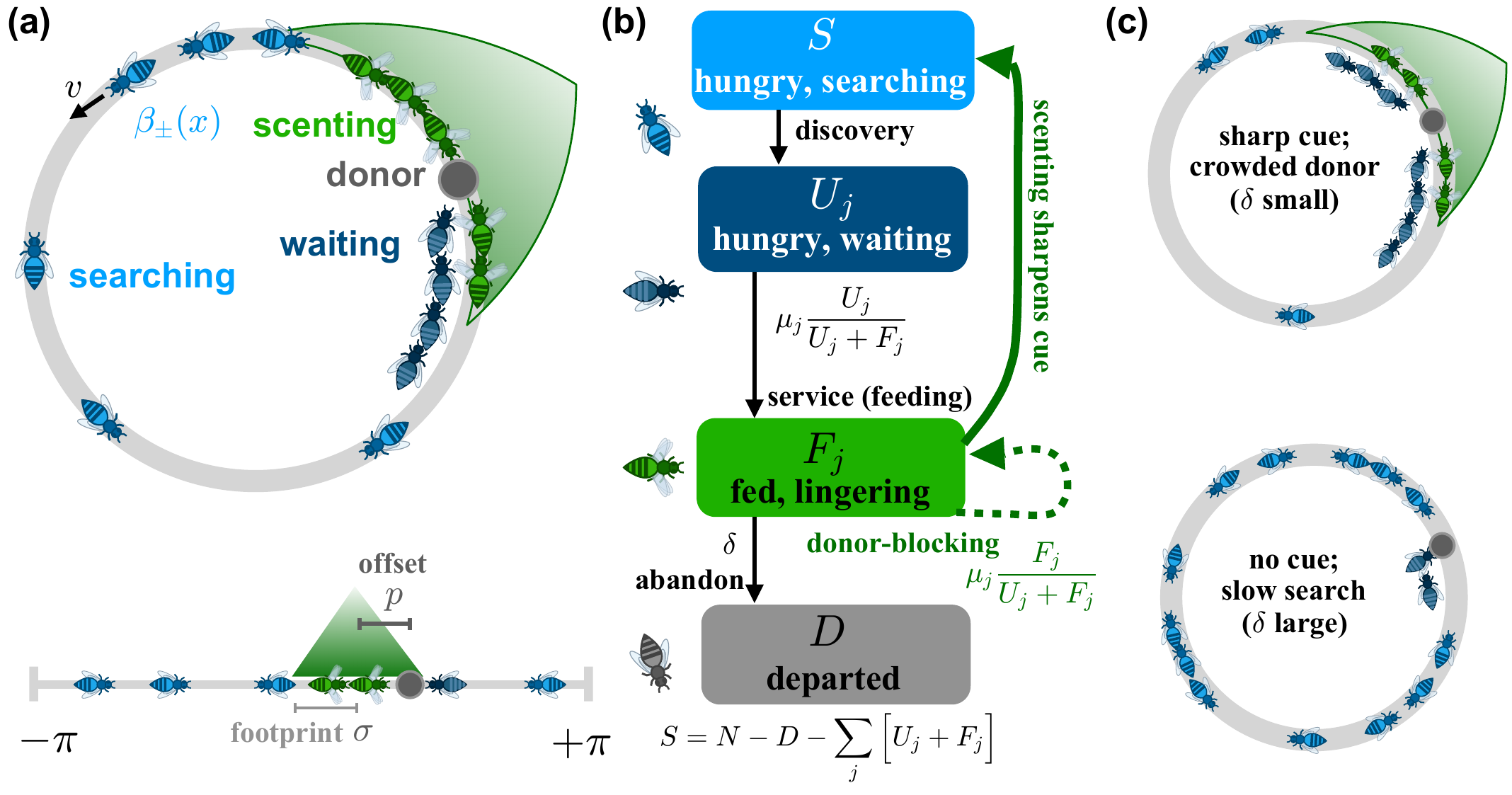}
\caption{\protect\justifying Stochastic search combined with random-order queueing for resources. {\bf (a)}~Search geometry. Bees are drawn heading in their direction of motion, and scenting bees face the donor and fan pheromone behind them, with the tent representing the time-averaged cue they sustain. Hungry bees move at speed $v$ on the ring $[-\pi,\pi)$ and reverse direction at rate $\beta_\pm(x)$ according to the local pheromone gradient. A donor absorbs searchers probabilistically with strength $\overline{\kappa}$. Scenting fed bees generate a tent cue of half-width $\sigma$ whose peak is displaced from the donor by a random offset $p$, whose spread narrows as the number of scenters grows. {\bf (b)}~Searchers $S$ discover a donor and join the hungry aggregation $U_j$. Service fires at rate $\mu_j$ and selects uniformly from $U_j+F_j$, so productive service occurs at rate $\nu_j=\mu_jU_j/(U_j+F_j)$, moving one bee from $U_j$ to $F_j$. Events selecting a fed bee transfer nothing (dashed arrow). Fed bees abandon at rate $\delta$ and enter $D$, while those remaining scent and generate the plume that raises the discovery rate of $S$ (green arrow). {\bf (c)}~At small $\delta$, lingering fed bees sharpen the cue but crowd the donor, making discovery fast and service slow. At large $\delta$, they leave too quickly for a coherent cue to form, so search is unaided while service is unimpeded.}
\label{fig:schematic}
\end{figure*}

In the model, hungry bees search by a velocity-jump process and encounter donors represented as partially absorbing targets, while fed bees linger near donors and contribute to recruitment cues (Fig.~\ref{fig:schematic}). Analysis of the single-donor search problem (Sec.~\ref{sec:search}) identifies when displaced cues remain useful. We then embed this search process in a conserved finite population (Sec.~\ref{sec:collective}), where searchers gather at donors, service selects randomly from the local crowd, and fed bees abandon at rate $\delta$. Feeding is fastest at an intermediate abandonment rate since rapid departure weakens recruitment, whereas slow departure congests the donor. With two donors of unequal service rate (Sec.~\ref{sec:two_donors}), the faster donor produces fed bees sooner and therefore recruits more strongly, directing bees toward the donor that can sustain greater throughput and relieving congestion at the slower queue.

\section{Model framework}
\label{sec:model}

\subsection{Search layer}
\label{sec:model_search}

We represent the searchable part of the hive by the ring $[-\pi,\pi)$ [Fig.~\ref{fig:schematic}(a)]. A hungry searcher moves at constant speed $v$, either right or left, and reverses direction as a Poisson process whose rate depends on the local recruitment cue. Donor $j$ lies at $x_j$ and is represented by an interaction region of half-width $a$ with absorption rate $\kappa$, or by its partially absorbing point-target limit $a\to0$ at fixed integrated strength $\overline\kappa=2\kappa a$, in which a searcher crossing the donor is captured with probability $q_0=1-e^{-\overline\kappa/v}$.

Bees scenting near donor $j$ generate a pheromone tent of half-width $\sigma$ centered at $x_j+p_j$, where $p_j$ is the displacement of the peak from the donor. We take $p_j$ to be random with zero mean, as when scenters surround the donor and face different directions, with a spread that narrows as the number of scenters grows. A scenting bee directs pheromone downstream of its position, so the peak need not lie over the bee or over the donor. A systematic displacement would enter as a nonzero mean of $p_j$, which we address in the Discussion. Only the local slope $c_j'(x)$ enters the search dynamics: it is $+1/\sigma$ on the rising flank $x_j+p_j-\sigma<x<x_j+p_j$, $-1/\sigma$ on the falling flank $x_j+p_j<x<x_j+p_j+\sigma$, and zero elsewhere, modulo $2\pi$. Searchers reverse out of state $\pm$ at rate
\begin{align}
\beta_\pm(x)=\beta_0\left[1\mp\frac{\gamma\,c_j'(x)}{s_m}\right],
\label{eq:turning_rate}
\end{align}
where $\beta_0$ is the unbiased reversal rate, $s_m$ a reference slope, and $\gamma$ the reliance on the cue. Thus, motion up the gradient is favored over motion down it, with positivity requiring $\gamma\le s_m\sigma$.

Writing $P_\pm(x,t)$ for the densities of right- and left-moving searchers gives
\begin{align}
\pd_t P_+ &= -v\pd_x P_+-\beta_+P_+ +\beta_-P_- -\kappa\chi_a P_+,\nonumber \\
\pd_t P_- &= +v\pd_x P_-+\beta_+P_+ -\beta_-P_- -\kappa\chi_a P_-,
\label{eq:single_stat}
\end{align}
with periodic boundary conditions, where $\chi_a(x)=1$ for $|x-x_j|\le a$ and zero otherwise. In the point-target limit, $\kappa\chi_a(x)\to\overline\kappa\,\delta(x-x_j)$. Searchers enter with no knowledge of donor position, so we use the uniform initial condition $P_\pm(x,0)=1/(4\pi)$.

\subsection{Service and abandonment at a donor}
\label{sec:model_service}

We model a fixed colony of $N$ bees as a continuous-time Markov chain with state $(S,U_j,F_j,D)$ together with the positions and headings of the $S$ searchers [Fig.~\ref{fig:schematic}(b)]. Here $S$ counts hungry bees searching in the bulk, $U_j$ those waiting at donor $j$, $F_j$ fed bees still at the donor, and $D$ fed bees that have departed. The population is conserved, $S+\sum_{j}(U_j+F_j)+D=N$. As the aggregation grows, fewer bees remain available to search, reducing the discovery rate. Donor $j$ serves at Poisson rate $\mu_j$, drawing one bee uniformly from the local aggregation at each event. Hungry bees at donor $j$ are fed at rate
\begin{align}
\nu_j(U_j,F_j)=\mu_j\frac{U_j}{U_j+F_j},
\label{eq:productive_service}
\end{align}
for $U_j+F_j>0$, where fed bees remaining at the donor block access by hungry bees, and each feeding event moves one bee from $U_j$ to $F_j$. Fed bees leave at rate $\delta$, setting the signaling-congestion tradeoff. Small $\delta$ strengthens the cue but crowds the donor, while large $\delta$ clears the aggregation and weakens the cue [Fig.~\ref{fig:schematic}(c)]. Thus the aggregation acts as a service queue for $U_j$ crowded by $F_j$, while the signaling fed bees generate the cue that raises the discovery rate of $S$, the green arrow in Fig.~\ref{fig:schematic}(b).

\subsection{Performance measure}
\label{sec:model_measure}

Food introduced through trophallaxis typically reaches only part of the colony over the relevant timescale~\cite{nixon1952}. We therefore measure performance by the time to feed a prescribed fraction $q$, consistent with colony-scale spreading measures based on the time to reach a fixed fraction of individuals~\cite{gernat_automated_2018}. Writing
\begin{align*}
N_f(t)=D(t)+\sum_j F_j(t)
\end{align*}
for the number fed by time $t$, we define
\begin{align}
T_q=\inf\left\{t\geq0\mathrel{}\middle|\mathrel{}N_f(t)\geq qN\right\}.
\label{eq:Tq_def}
\end{align}
Varying $q$ also resolves different parts of the tradeoff. Small $q$ weights the early part of a run, when few bees scent and the cue is weak, whereas large $q$ weights sustaining recruitment late in the run, when the searching pool is depleted. The variance of the half-colony time $T_{1/2}$ shrinks as $N$ grows, whereas $T_1$ is a highly variable extreme-value statistic set by the final arrivals.

\begin{figure*}[t!]
\centering
\includegraphics[width=\textwidth]{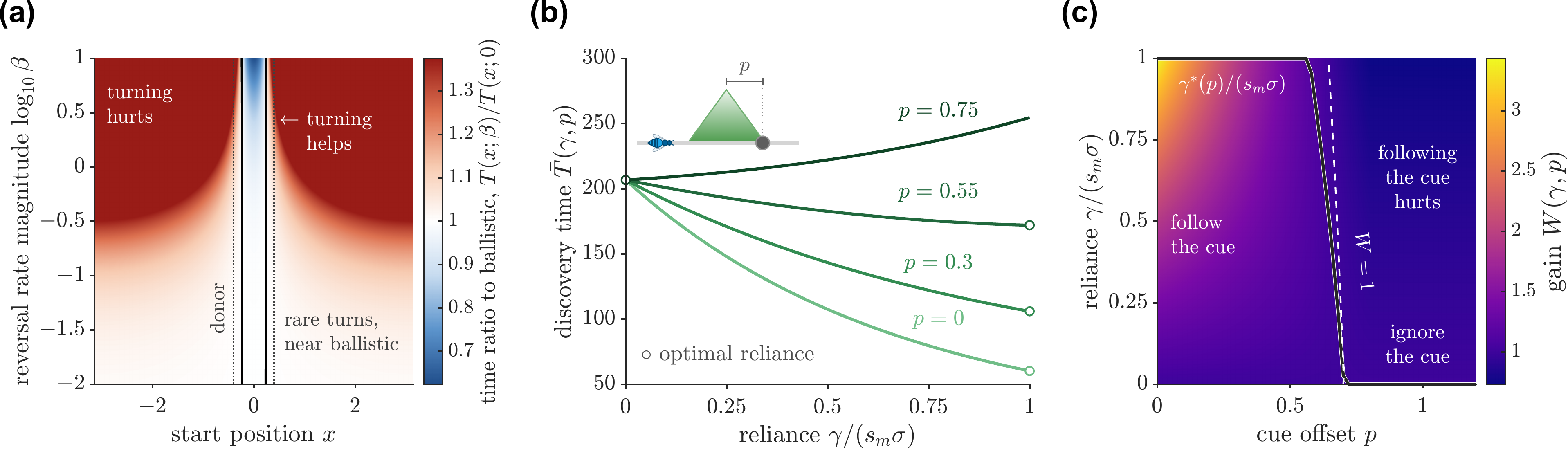}
\caption{\protect\justifying Centered cues speed discovery for searchers with well-tuned gradient-based turning strategies.
{\bf (a)}~Mean first-passage time relative to ballistic motion, $T(x;\beta)/T(x;0)$, for a donor of half-width $a=0.4$ and absorption rate $\kappa=1.2$, with $v=1$. Reversals help only near the donor center (blue). Dotted lines mark donor edges; the solid contour marks equal search times.
{\bf (b)}~Discovery time against normalized reliance for cue offsets $p=0$, $0.30$, $0.55$, and $0.75$ (light to dark), with minima marked. The inset illustrates the displaced cue.
{\bf (c)}~The gain $W(\gamma,p)=\bar T(0)/\bar T(\gamma,p)$ shows where following a cue helps ($W>1$) or hurts ($W<1$) search with a transition curve (dashed: $W=1$) that is sensitive to $p$. The solid curve marks the normalized reliance that maximizes $W$ for a given $p$. Other parameters as in Table~\ref{tab:params}.}
\label{fig:search}
\end{figure*}

We measure time in units of the mean service interval $1/\mu$ and length so that the ring has circumference $2\pi$, so a searcher with $v=1$ completes a circuit in $L=2\pi$. Four dimensionless parameters govern the search, besides the reliance $\gamma/s_m\sigma$ that a searcher chooses. The per-crossing capture probability $q_0$ sets how many circuits an unaided searcher needs. The reversals per circuit $\beta_0L$ set whether motion is ballistic or diffusive on the scale of the ring. The cue footprint $\sigma/\pi$ and the single-scenter scatter $p_0/\sigma$ set how much of the ring a plume covers and how reliably it points toward the donor. The colony adds four more: its size $N$, the fed fraction $q$, the traffic ratio $\rho=\lambda_0N/\mu$, which compares the colony's unaided discovery rate, $\lambda_0N$, with the donor's service rate $\mu$, so that $\rho>1$ means arrivals would outpace service, and the product $\delta L$, which compares the turnover of the scenting population with the circuit time. Recruitment matters most for a hard-to-find donor with a modest queue, $q_0\ll1$ and $\rho<1$, the parameter regime we study (Table~\ref{tab:params}). Appendix~\ref{app:methods} converts these dimensionless values to physical units for a typical assay.

\section{Discovery of a donor}
\label{sec:search}

We first quantify the rate at which a searcher discovers a donor, starting with a single donor and no cue, then adding a displaced cue that represents chemical signaling, and finally letting two donors compete for the same searcher. Figure~\ref{fig:search} summarizes the results.

\subsection{Turning slows unguided search}
\label{sec:single_target}

We begin with the unaided problem, which sets the baseline for recruitment. With $\gamma=0$ the turning rate \eqref{eq:turning_rate} is a constant $\beta$, and the mean first-passage time (MFPT) from each starting position and heading solves the backward equations associated with \eqref{eq:single_stat}~\cite{redner_guide_2001,gardiner_stochastic_2009,hillen_mfpt_2025,dorsogna_higher_2026}. Solved arc by arc and averaged over a uniform initial position and heading, it gives $\bar T(\beta)$ in closed form (Appendix~\ref{app:single_target}). Every term of $\bar T(\beta)$ is nondecreasing in $\beta$, so without a cue a searcher finds the donor most quickly when it never turns ($\beta=0$).

This optimum holds on average, not for every searcher. Figure~\ref{fig:search}(a) resolves the MFPT by starting position. Only a searcher that starts near the center of the donor's interaction region finds it faster at finite $\beta$, because reversals keep it within the absorbing zone instead of carrying it out and around the ring. Every other searcher is better off moving straight. A ballistic searcher is guaranteed to cross the donor again within one circuit, whereas each reversal sends it back over ground it has already covered. Averaged over starting positions this penalty dominates, consistent with the broader finding that, without information about where a target lies, searches built from long straight runs outperform frequent reorientation~\cite{viswanathan_optimizing_1999,benichou_intermittent_2011}. Turning becomes useful only once a signal points toward the donor.

For a point donor and a ballistic searcher the arrival statistics are explicit. The searcher crosses the donor once per circuit, and each crossing captures it independently with the per-crossing capture probability $q_0=1-e^{-\overline\kappa/v}$ of Sec.~\ref{sec:model_search}. It is therefore captured only at these crossings, after a geometrically distributed number of failed passes, and it is captured on crossing $n+1$ with probability $q_0(1-q_0)^{n}$. Averaging over the failed passes gives the unaided discovery rate
\begin{align}
\lambda_0=\frac{1}{\bar T_{\rm ball}}=\frac{v\,q_0}{\pi\left(2-q_0\right)}.
\label{eq:lambda0}
\end{align}
The mean search time counts $(1-q_0)/q_0$ unsuccessful loops plus the half loop a searcher travels on average before its first encounter. This rate sets the baseline for the collective model below, while the restriction of arrivals to discrete crossing times sets when this rate can be used to determine encounter statistics.

\subsection{Cues must envelop the donor to speed search}
\label{sec:chemo}

A pheromone cue biases turning, so that searchers reverse less often when heading up the gradient and more often when heading down it, speeding discovery of the donor. Because the tent is piecewise linear, the rates \eqref{eq:turning_rate} are piecewise constant, and the backward equations can be solved exactly on the arcs between the tent kinks and the donor. The response law enters only through the difference and sum of the two reversal rates, and a point donor enters only through matching conditions of strength $\eta=\tanh(\overline\kappa/2v)=q_0/(2-q_0)$ (Appendix~\ref{app:arcs}). We evaluate the resulting closed form numerically.

The averaged MFPT against reliance at several offsets [Fig.~\ref{fig:search}(b)] shows when a searcher should follow the cue and when it should ignore it. When the plume peak sits on the donor, the MFPT decreases monotonically in $\gamma$, and the optimal reliance is determined by the ceiling $s_m\sigma$. At larger offsets the optimal reliance falls to zero over a narrow range, beyond which ignoring the cue is optimal. To compare cues on a common scale we define the gain a searcher of reliance $\gamma$ obtains from a cue at offset $p$,
\begin{align}
W(\gamma,p)=\frac{\bar T(0)}{\bar T(\gamma,p)},
\label{eq:gain_surface}
\end{align}
the factor by which the cue speeds discovery relative to unaided search, so that $W(0,p)\equiv1$, with $W>1$ meaning the cue speeds discovery and $W<1$ meaning it slows it. The gain surface in Fig.~\ref{fig:search}(c) maps $W$ across the whole strategy plane of reliance and offset, with the optimal reliance $\gamma^\ast(p)$ tracing the structure seen in Fig.~\ref{fig:search}(b).

We define the critical offset $p^\ast$ as the offset at which a small reliance stops helping, $\pd_\gamma\bar T(\gamma,p)|_{\gamma=0}=0$. This transition is set by the width of the cue rather than by a balance of timescales. For $p\ge\sigma$ the donor lies outside the tent, both flanks push a biased searcher away from it, and so $p^\ast\le\sigma$ holds exactly. The computed threshold lies just below this bound and moves closer to it as the donor becomes more strongly absorbing (Appendix~\ref{app:arcs}, Fig.~\ref{fig:app_search}(b)). Once the peak lies more than $\sigma$ from the donor, following the cue is not merely uninformative but harmful, since any positive reliance lengthens search.

The best a searcher can do with a cue at offset $p$ is given by the attainable gain $G(p)=\max_\gamma W(\gamma,p)$. Following a cue never pays beyond the critical offset, so $G(p)=1$ for $p\ge p^\ast$. The largest gain occurs when the peak sits on the donor, and it is modest for our parameters (Table~\ref{tab:params}). This maximal gain is larger when the donor is harder to find and when searchers turn more often (Appendix~\ref{app:arcs}, Fig.~\ref{fig:app_search}(b)).

An intermediate reliance is optimal because of the time a searcher needs to travel from the plume peak back to the donor. A searcher that relies strongly on the cue stays close to the peak, so when the peak is offset from the donor by $p$ it must move down the rising flank, against the bias, to reach the donor. The mean time for this transit grows exponentially with the offset and with the imbalance between the two reversal rates on the flank (Appendix~\ref{app:ceiling}). Increasing $\gamma$ draws searchers toward the peak but also lengthens this transit, and which effect dominates depends on the offset.

\subsection{Competing donors split the searching population according to their discoverability}
\label{sec:two_targets}

We next study how competition impacts the speed and distribution of donor discovery, placing two point donors at $x_1$ and $x_2$ with strengths $\overline\kappa_j$. Let $\pi_j^\pm(x)$ be the probability of capture at donor $j$ first, starting from $x$ with heading $\pm$. Away from the donors these satisfy the homogeneous telegraph equations, and partial absorption has strength $\eta_j=\tanh(\overline\kappa_j/2v)$ (Appendix~\ref{app:two_targets}). The splitting probability has two clean limits (Appendix~\ref{app:two_targets}). A ballistic searcher encounters both donors many times before capture, each encounter an independent trial, so donor $j$ captures a fraction given by its relative absorptivity, $\eta_j/(\eta_1+\eta_2)$. A rapidly reversing searcher moves diffusively and is almost surely captured by the first donor it reaches, so the probability of reaching a donor decreases linearly with initial distance, evenly splitting the hitting probabilities over the ring.

\begin{figure*}[t!]
\centering
\includegraphics[width=\textwidth]{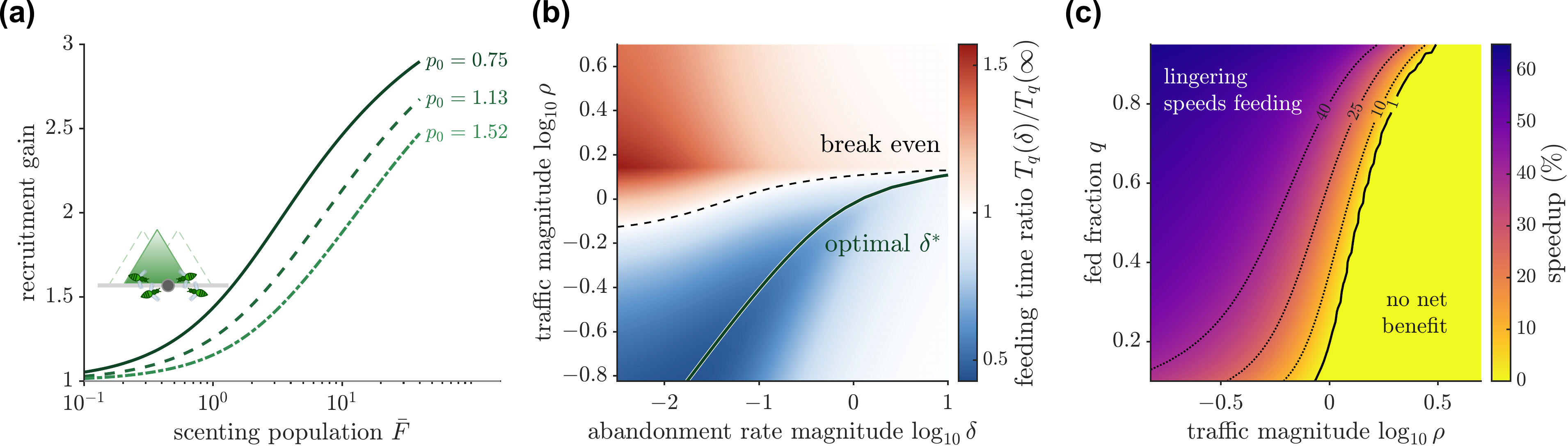}
\caption{\protect\justifying Lingering speeds feeding, and feeding is fastest at an intermediate abandonment rate. Times are in units of $\mu^{-1}$. {\bf (a)}~Recruitment gain $\widetilde G(\bar F)$ against scenting population with the offset kept stochastic, for $p_0=0.75$, $1.13$, $1.52$ (solid, dashed, dash-dot, darkest at the smallest offset, as in Fig.~\ref{fig:lingering2}(c)). The inset shows scenting bees fanning at a donor, producing a cue whose peak lands at a random offset (dashed outlines). {\bf (b)}~Feeding time relative to immediate departure against abandonment rate and traffic ratio $\rho=\lambda_0N/\mu$ at $q=1/2$, with the optimum $\delta^\ast(\rho)$ and the break-even contour. {\bf (c)}~Attainable speedup, $1-\min_\delta T_q(\delta)/T_q(\infty)$, against traffic ratio and fed fraction, contoured at $1$, $10$, $25$, and $40$ percent. The solid one-percent contour marks where lingering effectively stops helping, since the speedup falls continuously to zero across it.}
\label{fig:lingering}
\end{figure*}

Efficient search for donors is determined both by the fractions encountering each donor as well as the speed of those encounters. The Laplace transform $\hat g_j(x,s)$ of the first-passage density captured at donor $j$ can determine the encounter probability and time. At $s=0$ this returns the splitting probability, and one derivative gives the conditional time,
\begin{align}
\pi_j=\hat g_j(0),
\qquad
\pi_jT_j=-\left.\pd_s\hat g_j\right|_{s=0},
\label{eq:conditional_times}
\end{align}
with $T_j$ the MFPT conditioned on capture at $j$. A searcher reaches some donor at rate $1/\bar T$, and that donor is $j$ with probability $\pi_j$, so searchers discover donor $j$ at rate
\begin{align}
\Lambda_j=\frac{\pi_j}{\bar T},
\qquad
\bar T=\sum_k\pi_kT_k,
\label{eq:Lambda}
\end{align}
which is not $\pi_j/T_j$ unless the conditional times agree. Without cues they very nearly do at the reversal rate we use, even when the two absorptivities differ considerably (Appendix~\ref{app:two_targets}, Fig.~\ref{fig:app_search}(c)), so donors are discovered in proportion to their splitting probabilities.

Searchers cannot know the pheromone profiles ahead of time, so when cues surround both donors, we assume they apply a single reliance to both. We choose $\gamma$ to minimize the unconditional search time $\bar T$ and evaluate $\Lambda_j$ at that reliance. Because its reversal rates respond to the local gradient, the searcher still tracks cue profiles as they change over time. A cue mainly speeds discovery of its own donor (Appendix~\ref{app:two_cues}). If a cue is at one donor only, the advertised donor gains arrivals while the other loses a modest fraction of its own, so the cue speeds search far more than it redirects searchers. Those that still reach the unadvertised donor arrive sooner than they would without any cue, since they are the searchers not absorbed by the advertised donor. When both cues are present, the single-donor approximation $\Lambda_j(p_1,p_2)\approx\Lambda^0G(p_j)$ overestimates the arrival rates, although the ratio $\Lambda_1/\Lambda_2$ remains close to $G(p_1)/G(p_2)$ when both cues are centered over donors (Appendix~\ref{app:two_cues}). We therefore compute $\Lambda_j(p_1,p_2)$ by solving the first-passage problem with both cues present.

\section{Recruitment and congestion}
\label{sec:collective}

We now embed the search process in a conserved colony and ask how fed bees who remain at a donor accelerate its discovery, at the cost of crowding the donor they advertise. At a single donor, lingering strengthens recruitment but also increases competition for service, so an intermediate abandonment rate minimizes feeding time (Fig.~\ref{fig:lingering}). Throughout, the traffic ratio $\rho=\lambda_0N/\mu$ compares the rate at which the whole colony finds the donor with the rate at which it serves (Sec.~\ref{sec:model_measure}).

\subsection{Recruitment builds from one scenter}
\label{sec:gain}

Fed bees that remain at the donor release and fan scent, creating a collective cue whose reliability is most sensitive to fluctuations in their number when only a few bees are present. Hungry bees are fed at rate $\nu$, and each fed bee leaves independently at rate $\delta$, so the number of fed bees at the donor is Poisson with mean $\bar F=\nu/\delta$~\cite{kleinrock_queueing_1975}, once there is a steady stream of hungry bees. Each scenter’s position is drawn from a normal distribution centered on the donor with variance $p_0^2$. We place the peak of the plume built by $k$ of them at the mean of those positions, as a smooth plume would, so the peak offset has zero mean and variance $p_0^2/k$. A literal sum of tent cues sharpens more slowly, which changes the size of the benefit at large scenting populations but not its sign or the absence of a minimum scenter number (Appendix~\ref{app:offset}). Discovery at a fixed cue is nearly Poisson, but the cue changes constantly, so we estimate the discovery rate by averaging over cue configurations and let each searcher use a single reliance $\gamma$, since it cannot observe where the plume peak lies. The recruitment gain is then
\begin{align}
\widetilde G(\bar F)&=\max_{\gamma}\ \sum_{k\geq0}e^{-\bar F}\frac{\bar F^{k}}{k!}\,a_k(\gamma),\label{eq:averaged_gain}
\end{align}
where \mbox{$a_k(\gamma)=\int W(\gamma,p)\,\mathcal N(p;0,p_0^2/k)\,\mathrm{d}p$} is the $k$-scenter gain, the average of $W$ over the peak offset at fixed reliance, and $a_0\equiv1$ since no scenters means no plume. The optimal strategy is the maximizer $\gamma$ after averaging (Appendix~\ref{app:gain}).

The averaged gain increases even with one scenter [Fig.~\ref{fig:lingering}(a)], since even for small $\bar F$, some realizations of the offset land within $p^\ast$ of the donor and benefit searchers. The slope at the origin is therefore
\begin{align}
\widetilde G'(0^+)=g_1-1,
\qquad
g_1=\max_\gamma a_1(\gamma),
\label{eq:g1}
\end{align}
indicating the reliance that gains the most from a single scenter. The first scenter helps when it lands within $p^\ast$ of the donor and is otherwise misleading, so $g_1$ falls to one as $p_0$ grows, beyond which at least two scenters are needed to speed search (Appendix~\ref{app:gain}, Fig.~\ref{fig:app_gain}(a)). The gain then saturates slowly toward its ceiling.

Approximating recruitment as a rate needs justification, since arrivals at a donor are first passages of individual searchers and not a Poisson process. Turning spreads the discrete crossing times of a ballistic searcher out in time, so at $\beta_0=1$ the discovery time is close to an exponential waiting time (Appendix~\ref{app:gap}). The cue also turns over far faster than a search lasts (Appendix~\ref{app:gain}), so we approximate the arrival rate at one donor as
\begin{align}
A\approx S\,\lambda_0\,\widetilde G(F).
\label{eq:reduced_arrival}
\end{align}
Each of the $S$ searching bees finds the donor at its unaided rate $\lambda_0$ multiplied by the recruitment gain $\widetilde G(F)$ due to the $F$ scenting bees. A cue guides search only if it persists longer than a circuit $L=2\pi/v$, the time searchers need to adjust, and the scenting population must turn over many times within one search so that searchers sample many cue configurations. Together these give the valid approximation regime
\begin{align}
\frac{2\pi}{v}\ \ll\ \frac{1}{\delta}\ \ll\ \frac{1}{\lambda_0},
\label{eq:coherence_window}
\end{align}
so lingering speeds feeding only when the donor is hard to find. The circuit time estimates the relaxation time of the biased search, which its spectral gap gives precisely in any geometry (Appendix~\ref{app:gap}).

\subsection{Lingering generates congestion without a capacity limit}
\label{sec:congestion}

Lingering also slows service as those signaling block hungry bees waiting at the donor. Inside the range \eqref{eq:coherence_window}, the discovery rate $A$ changes slowly compared with the relaxation of the aggregation, so the flows through the donor balance. Hungry bees join at rate $A$, the donor feeds them at rate $\nu=\mu U/(U+F)$ since a service event reaches a hungry bee with probability $U/(U+F)$, and each fed bee leaves at rate $\delta$. Balancing arrivals against service, $A=\nu$, and service against abandonment, $\nu=\delta F$, gives the quasi-stationary aggregation sizes
\begin{align}
F^\ast=\frac{A}{\delta},
\qquad
U^\ast=\frac{A^2}{\delta(\mu-A)},
\label{eq:qss_crowd}
\end{align}
These sizes exist only for $A<\mu$, since the feeding rate $\nu$ approaches $\mu$ but never reaches it, so faster arrivals leave a queue that grows instead of settling. They also fail once $\delta$ falls to the order of the discovery rate $\lambda_0$, where the aggregation relaxes more slowly than $A$ changes. Within their range, they expose two features of the tradeoff. The fraction of service events that lead to a hungry bee being fed, $U^\ast/(U^\ast+F^\ast)=A/\mu$, does not depend on $\delta$, so lingering changes how many bees stand at the donor $F$ but not the fraction of service that is wasted. Bees at the donor also cannot search, so a larger aggregation reduces $A$ through Eq.~\eqref{eq:reduced_arrival}. Congestion therefore slows feeding even though we impose no capacity limit at the donor. A hungry bee waits a mean time $U^\ast/A=A/[\delta(\mu-A)]$, which lengthens as $\delta$ falls because fed bees that stay crowd out service, whereas the recruitment benefit is bounded by $G(0)$.

\begin{figure*}[t!]
\centering
\includegraphics[width=\textwidth]{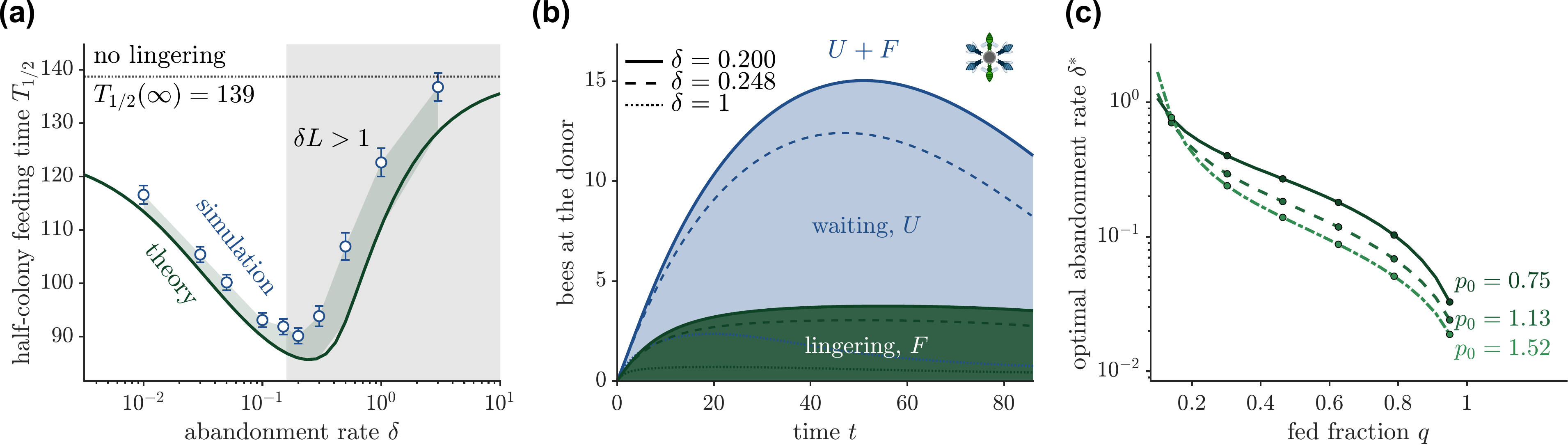}
\caption{\protect\justifying The optimal abandonment rate falls as a larger fraction of the colony must be fed, and simulation supports the reduction near the optimum. Times are in units of $\mu^{-1}$. {\bf (a)}~The reduction \eqref{eq:mean_field} (curve) against simulation of the individual-based model (markers, $200$ realizations, bars $\pm1.96$ standard errors) at $\rho=0.6$, $q=1/2$. Shading marks the part of the predicted speedup the simulations do not achieve. In the gray band, $\delta L>1$, fed bees leave faster than a searcher can circle the ring. The dotted line is the median feeding time $T_{1/2}$ when fed bees depart immediately ($\delta \to \infty$). {\bf (b)}~Fed bees $F$ and total occupancy $U+F$ at the donor over time from the reduced dynamics at $\delta=0.20$ (simulated optimum, shaded), $0.248$ (reduced optimum), and $1.0$. The inset shows scenting and waiting bees mixed at the donor. {\bf (c)}~Optimal abandonment rate against fed fraction at $\rho=0.6$.}
\label{fig:lingering2}
\end{figure*}

To find when lingering speeds feeding, we write the time $T_q$ to feed a fraction $q$ of the colony as an integral over the searching population and expand it for large $\delta$ using $\widetilde G(F)=1+(g_1-1)F$ (Appendix~\ref{app:limits}, Fig.~\ref{fig:app_limits}(a)), which gives
\begin{align}
T_q(\delta)=\frac{\ell_{q}}{\lambda_0}
+\frac{1}{\delta}\left[\frac{A_{q}}{\mu-A_{q}}-\left(g_1-1\right)qN\right]
+O\!\left(\delta^{-2}\right),
\label{eq:large_delta}
\end{align}
where $\ell_q=\ln[1/(1-q)]$ and $A_q=\lambda_0(1-q)N$ is the discovery rate at the end of the run, when the target fraction $q$ of the colony has been fed and about $(1-q)N$ searchers remain. The first term is the unaided feeding time, exact because $S$ decays exponentially at rate $\lambda_0$ when there is no cue. Both corrections scale with $1/\delta$, the mean time a fed bee stays. The benefit is the time saved because each of the $qN$ bees fed during the run, while present, raises the discovery rate of the remaining searchers by the fraction $g_1-1$. The cost is the queue standing at the donor at the end of the run, which must also be fed before the fraction $q$ is reached, and it grows without bound as the load $A_q/\mu$ approaches one. Lingering therefore speeds feeding exactly when the recruitment benefit of an additional fed bee exceeds the congestion it creates, that is, when the bracketed term is negative,
\begin{align}
\left(g_1-1\right)qN\left(\mu-A_{q}\right)>A_{q}.
\label{eq:linger_criterion}
\end{align}
This requires spare service capacity at the end of the run, $\rho<1/(1-q)$ with $\rho=\lambda_0N/\mu$, and becomes easier to satisfy at larger $q$, since the benefit grows with the target fraction $q$, while the cost falls as fewer searchers remain to arrive at the end of the run. No quorum of scenters enters, since the cue geometry enters only through $g_1$. At the other extreme, indefinite lingering is costly, since fed bees that never leave dilute service for the remainder of the run, and it stays well above the best feeding time at every fed fraction we examined (Appendix~\ref{app:limits}, Fig.~\ref{fig:app_limits}(a)). Feeding is thus fastest at an intermediate abandonment rate. Near that optimum, recruitment drives the load $A/\mu$ toward one and Eq.~\eqref{eq:qss_crowd} loses accuracy, so we determine $\delta^\ast$ with the full reduced dynamics and with simulations [Fig.~\ref{fig:lingering}(b) and Fig.~\ref{fig:lingering2}(a,c)].

\subsection{Feeding is fastest at an intermediate abandonment rate}
\label{sec:single_donor}

For one donor, the reduced dynamics are
\begin{align}
\dot S&=-A,
&\dot U&=A-\mu\frac{U}{U+F},\nonumber\\
\dot F&=\mu\frac{U}{U+F}-\delta F,
&\dot D&=\delta F,
\label{eq:mean_field}
\end{align}
with $A=S\lambda_0\widetilde G(F)$. Under random-order service, the only closure beyond Eq.~\eqref{eq:reduced_arrival} replaces $\langle U/(U+F)\rangle$ by its mean-field value, which is accurate for the small number of bees gathered at the donor near the optimum. We call Eq.~\eqref{eq:mean_field} the reduction and compare it with individual-based simulations that track every searcher's position and heading and every scenter's mark, using the parameters in Table~\ref{tab:params}. For immediate departure, there is no cue or congestion and the reduced dynamics give the unaided feeding time $\ell_q/\lambda_0$ exactly.

In simulations, letting fed bees linger substantially shortens the time to feed half the colony [Fig.~\ref{fig:lingering2}(a)]. The reduction agrees closely with the simulations inside the range \eqref{eq:coherence_window} and overestimates the benefit beyond it, where the cue turns over too quickly for searchers to follow. Because the simulated optimum lies just past the upper edge of that range, the reduction places $\delta^\ast$ somewhat too high (Appendix~\ref{app:gap}, Fig.~\ref{fig:app_limits}(b)).

The aggregation passes through three phases over a run [Fig.~\ref{fig:lingering2}(b)]. Few bees have been fed at first, the cue is weak, and searchers arrive at close to the unaided rate. As fed bees accumulate at the donor, the cue sharpens and arrivals outpace service, so hungry bees gather at the donor and the load $A/\mu$ briefly approaches one at the optimum. Late in the run the searching population is reduced, arrivals slow, and the queue drains while the cue persists. Congestion slows feeding by lowering the rate at which an arriving hungry bee is fed. At the optimum this effect is mild, since only a few fed bees remain at the donor and most of those present are hungry bees waiting. The optimum is also insensitive to the service-time distribution, since many exchanges are needed to feed half the colony and fluctuations in individual service times average out (Appendix~\ref{app:service}, Fig.~\ref{fig:app_limits}(c)).

The optimal abandonment rate depends strongly on the fraction $q$ of the colony that must be fed [Fig.~\ref{fig:lingering2}(c)]. As $q$ increases, $\delta^\ast$ falls steeply while the speedup grows. The cue is generated by fed bees and weakens late in the feeding process, so for larger $q$ sustaining the signal matters more than avoiding congestion early, and fed bees should linger longer.

\begin{figure*}[t!]
\centering
\includegraphics[width=\textwidth]{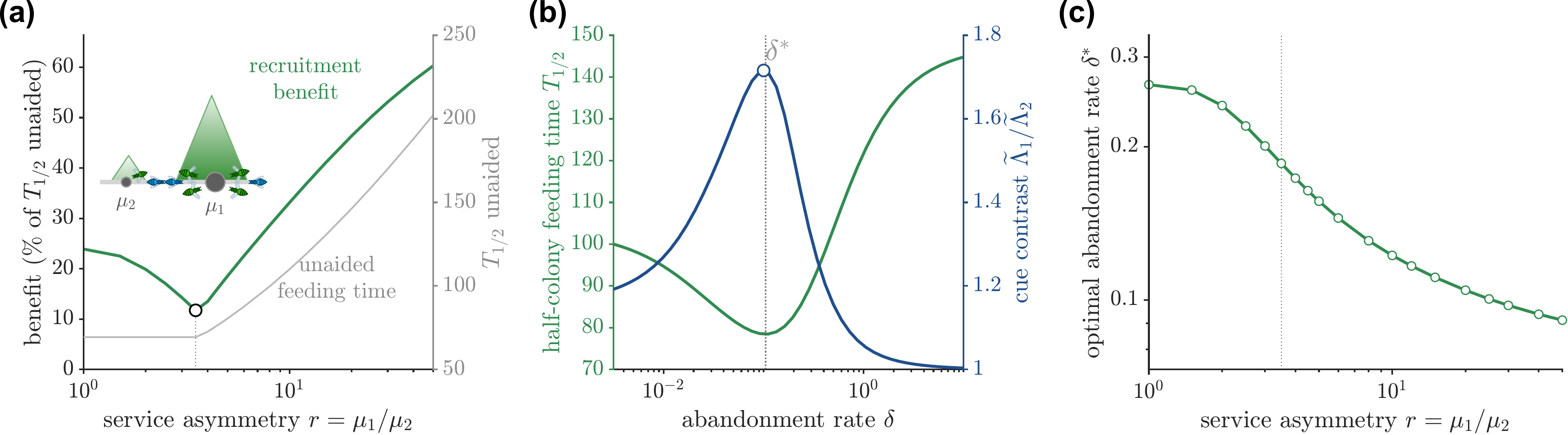}
\caption{\protect\justifying A cue that reports throughput steers searchers toward the faster donor and sets the optimal residence time. Two donors share one abandonment rate, with total capacity $\mu_1+\mu_2=2$ fixed and asymmetry $r=\mu_1/\mu_2$ varied. {\bf (a)}~Recruitment benefit as a percentage of the unaided feeding time (green, left axis), with the unaided feeding time itself in gray (right axis). The open marker indicates the minimum benefit. The inset depicts a fast donor ($\mu_1$) with a strong cue drawing searchers away from a slow donor ($\mu_2$) with a faint one. {\bf (b)}~Half-colony feeding time (green, left axis) and cue contrast $\widetilde\Lambda_1/\widetilde\Lambda_2$ (blue, right axis) against $\delta$ at $r=20$. The dotted line marks the feeding-time optimum $\delta^\ast$ and the marker the contrast maximum, which nearly coincide. {\bf (c)}~Optimal abandonment rate as a function of donor asymmetry. The dotted line marks the minimum benefit in (a).}
\label{fig:twodonor}
\end{figure*}

As the traffic ratio $\rho=\lambda_0N/\mu$ increases, the optimal abandonment rate rises steeply and the speedup decreases [Fig.~\ref{fig:lingering}(b)]. The boundary of the lingering regime moves to higher traffic ratios as $q$ increases [Fig.~\ref{fig:lingering}(c)], since the recruitment benefit accumulates over the run while the congestion cost stays roughly constant, as the criterion \eqref{eq:linger_criterion} anticipates. Because $\widetilde G'(0^+)>0$, the speedup vanishes continuously. The model also predicts that a more discoverable donor raises $\lambda_0$ and increases $\delta^\ast$, so that easily found donors should be abandoned sooner, and that larger colonies served at the same rate should show shorter residence times after feeding.

\section{Competing donors}
\label{sec:two_donors}

We now consider how the colony should search for two donors with unequal service rates and ask how recruitment distributes searchers between them. Our results are summarized in Fig.~\ref{fig:twodonor}.

\subsection{Recruitment cues scale with donor throughput}
\label{sec:throughput}

With two donors, the state is $(S,U_1,F_1,U_2,F_2,D)$, conserved as in Sec.~\ref{sec:model_service}. Each donor serves in random order at rate $\mu_j$, fed bees leave both donors at the same rate $\delta$, and searchers arrive at donor $j$ at rate $A_j=S\widetilde\Lambda_j(F_1,F_2)$. Here $\widetilde\Lambda_j$ is the two-cue analog of Eq.~\eqref{eq:averaged_gain}, averaged over the scenting populations and offsets at both donors, with a single reliance chosen to maximize the total rate and with offsets beyond the cue half-width treated as no cue (Appendix~\ref{app:two_cues}). Both donors share the abandonment rate $\delta$ and neither has a capacity limit, so differences in recruitment arise only from the dynamics of arrivals and departures. Applying the quasi-stationary balance of Sec.~\ref{sec:congestion} at each donor gives $F_j^\ast=\nu_j/\delta$, so the cue at donor $j$ is set by its throughput $\nu_j$. When donor $j$ keeps up with arrivals, $\nu_j$ equals its arrival rate $A_j$, and when a queue forms, $\nu_j$ approaches the service rate $\mu_j$.

With total service capacity $\mu_1+\mu_2$ held fixed, the recruitment benefit varies nonmonotonically with donor asymmetry $r=\mu_1/\mu_2$ [Fig.~\ref{fig:twodonor}(a)]. For small asymmetries, it decreases, reaching a minimum when the slower donor receives more searchers than it can serve, and then increases for intermediate asymmetries as the cue becomes strong enough to redirect searchers toward the faster donor. This increase in recruitment benefit also partly reflects an increase in feeding time without recruitment. In the absence of recruitment, searchers divide between donors according to how readily they encounter each one, rather than how quickly each donor can serve them. As the slower donor’s capacity falls, it therefore continues to attract the same share of searchers but takes longer to serve them. Recruitment reduces this mismatch by redirecting searchers toward the faster donor (Sec.~\ref{sec:two_targets}).

\begin{figure*}[t!]
\centering
\includegraphics[width=\textwidth]{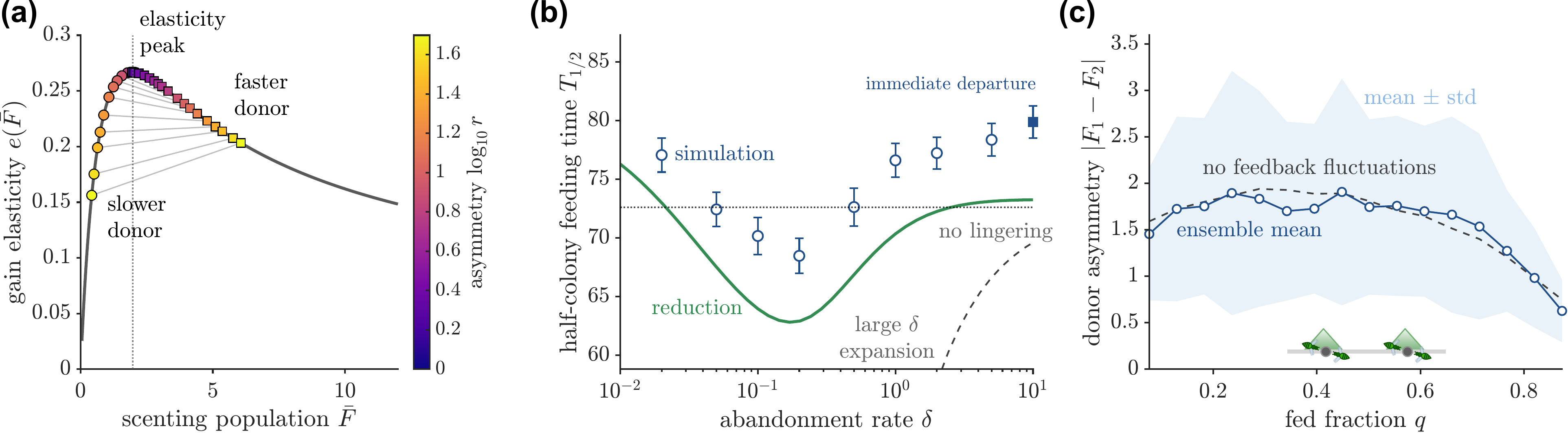}
\caption{\protect\justifying Fed populations follow the gain elasticity, the reduction tracks simulation up to a constant offset, and identical donors do not break symmetry beyond fluctuations. {\bf (a)}~Gain elasticity $e(\bar F)$ with the fed populations at the optimum, slower donor (circles) and faster donor (squares), joined at equal $r$ and colored by asymmetry. {\bf (b)}~Reduction and its large-$\delta$ expansion against simulation of the individual-based model at $r=4$ ($200$ realizations, bars $\pm1.96$ standard errors). The dotted line is the feeding time of a colony in which fed bees depart immediately, and the filled square its simulated counterpart. {\bf (c)}~Asymmetry $|F_1-F_2|$ at identical donors against the fed fraction $q$, ensemble mean over $120$ realizations with a $\pm1$ standard deviation band, against the no-feedback scale $\sqrt{4\bar F/\pi}$ (dashed). The inset depicts the two identical donors. Parameters as in Fig.~\ref{fig:lingering}, with $q=1/2$, except in (c), which runs to $q=0.9$.}
\label{fig:twodonor2}
\end{figure*}

\subsection{Allocation determines the optimal abandonment rate}
\label{sec:contrast}

Directing searchers toward the faster donor requires a difference in cue strength between the donors. Changing the abandonment rate rescales the fed populations at both donors by the same factor, so the contrast between the cues depends on how strongly each gain responds to a proportional change in its fed population. This sensitivity is the elasticity of the gain,
\begin{align}
e(F)\equiv\frac{\mathrm{d}\log\widetilde G(F)}{\mathrm{d}\log F}=\frac{F\,\widetilde G'(F)}{\widetilde G(F)}.
\label{eq:gain_elasticity}
\end{align}
Unlike the slope $\widetilde G'(F)$, the elasticity has no units and does not depend on the size of the population, so it can be compared directly between donors with different numbers of fed bees. For example, when $e(F)=0.25$, a one-percent increase in the fed population produces approximately a quarter-percent increase in recruitment gain. Near saturation, additional fed bees contribute little to the gain and the elasticity approaches zero. Using $F_j=\nu_j/\delta$ at fixed throughput, the product approximation gives
\begin{align}
\frac{\mathrm{d}\log(\widetilde\Lambda_1/\widetilde\Lambda_2)}{\mathrm{d}\log\delta}=e(F_2)-e(F_1).
\end{align}
Raising the abandonment rate shrinks both fed populations by the same factor, and each gain falls by its elasticity times that percentage, so the cue with the larger elasticity loses proportionally more. The contrast between the cues therefore grows with $\delta$ when $e(F_2)>e(F_1)$, shrinks when $e(F_1)>e(F_2)$, and is stationary with respect to $\delta$ when the two gains change by the same percentage, or equivalently when their elasticities are equal, $e(F_1)=e(F_2)$. Since the elasticity is single peaked (Sec.~\ref{sec:gain}), this condition places the two fed populations on opposite sides of its peak, and as the asymmetry grows the population at the faster donor moves onto the saturating side while that at the slower donor moves down the rising side [Fig.~\ref{fig:twodonor2}(a)]. Their elasticities are not exactly equal, since the product approximation does not capture the two-cue rates precisely. If fed bees linger longer, both populations move onto the saturating side, where both cues approach their maximum and the difference between them shrinks. The optimal abandonment rate is therefore set by the need to preserve contrast between the cues, and consistent with this, at strong asymmetry the feeding-time minimum and the contrast maximum nearly coincide [Fig.~\ref{fig:twodonor}(b)].

The optimal abandonment rate $\delta^\ast$ changes only modestly as donor asymmetry grows [Fig.~\ref{fig:twodonor}(c)]. At strong asymmetry, the faster donor accumulates scenters while the slower donor is left with almost none. The optimal rate thus varies far less than the asymmetry itself. This weak dependence matters because individual bees in our model have no way to measure the asymmetry between donors directly.

The reduction captures the dependence of feeding time on abandonment rate, with a small, nearly constant underestimate relative to the individual-based simulations [Fig.~\ref{fig:twodonor2}(b)]. A similar offset persists when fed bees depart immediately, indicating that it reflects the unaided search approximation rather than the dynamics of lingering (Appendix~\ref{app:limits}).

\subsection{Recruitment does not break symmetry between identical donors}
\label{sec:symmetry}

More fed bees at a donor sharpen its cue, which draws more searchers and produces more fed bees, so recruitment provides positive feedback, which could favor one donor over the other even when they are identical. Linearizing the reduced dynamics about the symmetric quasi-stationary state, the antisymmetric perturbation has a negative trace and grows only if the net rate at which a fed bee recruits replacements at its own donor exceeds its departure rate $\delta$. At the quasi-stationary state this condition is
\begin{align}
\frac{\pd\log\widetilde\Lambda_1}{\pd\log F_1}
-\frac{\pd\log\widetilde\Lambda_1}{\pd\log F_2}>1,
\label{eq:symmetry_criterion}
\end{align}
independent of $\mu$, $N$, and $\delta$, because the ratio of those two rates is a difference of log-elasticities. A fed bee must therefore recruit more than one replacement before it leaves. The elasticity of a donor's arrival rate to its own fed population stays well below one and the cross term is small and mostly negative, leaving the feedback well short of this threshold.

When the donors have equal service rates, the difference in their fed populations, $|F_1-F_2|$, remains comparable to the fluctuations expected for independent Poisson populations without feedback, for which the expected $|F_1-F_2|$ is $\sqrt{4\bar F/\pi}$, and no simulated run ends with a large imbalance [Fig.~\ref{fig:twodonor2}(c)]. Averaging over additional scenters improves cue precision but yields diminishing gains, keeping the recruitment feedback below the threshold for symmetry breaking. Recruitment thus directs searchers toward faster donors without locking the colony onto one of two equally productive donors.

\section{Discussion}
\label{sec:discussion}

Social insect colonies must distribute food without any individual tracking global supply or demand~\cite{greenwald_ant_2015,greenwald_individual_2018,meurville_trophallaxis_2021}. Models of trophallaxis usually describe a network of pairwise exchanges~\cite{gharooni-fard_data-driven_2020,greenwald_colony_2019}. Our model connects these exchanges to the search that precedes them. Fed bees become information producers whose presence helps nestmates find food but also crowds the donor. Their residence therefore controls both the value and the cost of recruitment, making departure after feeding part of the colony's food-distribution strategy.

The benefit of recruitment can grow smoothly even when individual bees scent only upon detecting the scent of others. Honeybees searching for their queen behave this way, with each scenter amplifying the signal of those before it~\cite{nguyen_flow-mediated_2021,nguyen_shortest_2023}. In our model, averaging over fluctuations in scenter number and position removes the minimum number of scenters that a description based on a typical cue would require, since even a single scenter sometimes lands close enough to the donor to help.

Recruitment also changes the role of congestion. Nectar foragers that wait long for a receiver perform tremble dances that recruit unloaders and suppress further foraging~\cite{seeley_social_1989,seeley_tremble_1992,anderson_worker_1999,anderson_task_1999,ratnieks_task_1999}, while ants at crowded food sources deposit less trail pheromone~\cite{czaczkes_negative_2012,czaczkes_how_2014}. Abandonment and inhibitory signals can thus restore flexibility lost to recruitment~\cite{bidari_social_2019}. Here, fed bees at the donor continue to attract searchers, so the same aggregation supplies information and creates congestion. The optimal residence balances these effects and depends on how much of the colony must be fed and how difficult the donor is to find.

Residence also sets the lifetime of the information. Trail pheromones must persist long enough to organize traffic but decay quickly enough to permit a response to change~\cite{doiron_balancing_2010}. In our model, cue persistence is controlled by the turnover of the bees producing it, and recruitment becomes less effective when that turnover outpaces the searchers' response. A similar constraint may apply to scenting networks in swarms, where each scenter must remain long enough for bees downstream to respond~\cite{nguyen_flow-mediated_2021}.

Because only fed bees scent in our model, the cue reports donor throughput and guides searchers toward faster donors without requiring any bee to compare them directly. This mechanism is consistent with observations that honeybees expose the Nasonov gland more readily at more profitable food sources~\cite{free_role_1972,fernandez_changes_2001,fernandez_reward_2003}. It also permits redistribution without persistent preference between equally productive donors. Negative feedback is thought to prevent such symmetry breaking and preserve flexibility in ant foraging~\cite{czaczkes_negative_2012,czaczkes_how_2014}; here, diminishing gains from additional scenters limit amplification even though the cue is purely attractive.

A fed bee bears the cost of lingering after feeding, while providing useful positioning information to its nestmates. The abandonment rate that minimizes colony feeding time is therefore a collective optimum, an instance of the distinction between individually and collectively adaptive foraging strategies~\cite{garg_individual_2022}. The model identifies when information production is valuable to the colony but does not specify how individual bees would adopt the corresponding residence times.

Some model limitations suggest extensions that could yield a richer set of experimental predictions. Effective arrival rates overestimate recruitment when cues turn over faster than searchers can respond, and the two-donor approximation treats strongly displaced cues as absent rather than harmful. We also take the cue peak to be unbiased about the donor. If scenters share a common orientation, so that plumes are displaced systematically downstream, $p_j$ acquires a nonzero mean, and recruitment helps only while that mean lies within the critical offset $p^\ast$ of Sec.~\ref{sec:chemo}. The symmetric tent cue also omits the directional pheromone flow generated by fanning bees~\cite{nguyen_flow-mediated_2021}. Fed bees scent at a constant rate and share one abandonment rate, whereas bees in dense clusters may stop scenting~\cite{nguyen_flow-mediated_2021} or adjust their residence to local conditions. Extending the model from stationary donors on a one-dimensional ring to moving bees and donors in two dimensions would test how geometry changes the balance between cue persistence, discovery, and congestion. Search times depend on geometry, since a ring forces repeated passes over the donor, so we treat the unaided discovery rate as a parameter.

Experiments could test this balance by tracking feeding, scenting, and departure together. Barcoded tracking resolves trophallactic events in bees and ants~\cite{gernat_automated_2018,greenwald_ant_2015}; combined with arena assays of collective scenting~\cite{nguyen_flow-mediated_2021,nguyen_shortest_2023}, it could measure how long fed bees remain near donors and how their presence affects subsequent arrivals. Varying donor accessibility and the fraction of the colony requiring food would test the predicted adjustment of residence times. Fed bees should leave easily found donors sooner and linger longer when more of the colony must be fed, while donors with higher delivery rates should attract a larger share of searchers.

In honeybee colonies, a fed bee can help its nestmates find the donor by remaining nearby and scenting. Its residence sustains the recruitment cue but also contributes to crowding, linking the success of one food exchange to the speed of those that follow.

Overall, our mathematical model formalizes the complex dynamics of distributed food-sharing networks while coupling the search dynamics to food dissemination. We make several testable predictions, for example, that searchers find a donor fastest when the recruitment cue is centered on it rather than displaced to one side, and that fed bees should linger longer after feeding when more of the colony is hungry. The same tradeoff between recruitment and congestion arises wherever served individuals both advertise and occupy a resource across a wide range of living and engineered systems: from neutrophils relaying chemokines after reaching a wound, to drivers reporting an open charging station they are using and robots beaconing a found target while blocking the rest of the swarm from reaching it.

\begin{acknowledgments}
We are grateful to Richard Terrile for helpful discussions.
ZPK acknowledges support from the National Institute for Theory and Mathematics in Biology through the National Science Foundation (grant number DMS-2235451) and the Simons Foundation (grant number MP-TMPS-00005320). This work was also supported by the National Science Foundation under grants DMS-2527338 (ZPK, OP) and IIS-2616531 (ZPK), and CPS-2433428 (OP).
\end{acknowledgments}

\section*{Author contributions}
WM contributed methodology, software, formal analysis, investigation, original drafting, and manuscript revision. OP contributed supervision and manuscript revision. ZPK contributed conceptualization, methodology, supervision, funding acquisition, original drafting, and manuscript revision.

\section*{Data availability}
No experimental data were used. The MATLAB code and parameters that regenerate every figure are publicly available under an MIT license~\cite{magrogan_trophallaxis_code_2026}.

\appendix

\section{Parameters and numerical methods}
\label{app:methods}

\begin{table}[b]
\caption{\protect\justifying Representative parameters. Every result uses these values unless a figure states otherwise. Values are dimensionless, with time in units of $1/\mu$ and length in units of $v/\mu$. The cited results anchor the parameter choices.}
\label{tab:params}
\begin{ruledtabular}
\begin{tabular}{llc}
symbol & meaning & value\\
\hline
$N$ & colony size~\cite{gharooni-fard_data-driven_2020} & $120$\\
$\mu$ & service rate per donor~\cite{gernat_automated_2018,choi_individual_2020,gharooni-fard_data-driven_2020} & $1$\\
$v$ & searcher speed~\cite{gharooni-fard_data-driven_2020} & $1$\\
$\beta_0$ & unbiased reversal rate & $1$\\
$\sigma$ & cue half-width~\cite{nguyen_flow-mediated_2021,gharooni-fard_data-driven_2020} & $0.8$\\
$s_m$ & reference slope & $1$\\
$\overline\kappa$ & integrated donor absorptivity & $0.0314$\\
$p_0$ & single-scenter offset spread~\cite{nguyen_flow-mediated_2021} & $0.75$\\
$\lambda_0$ & unaided discovery rate, Eq.~\eqref{eq:lambda0} & $0.005$\\
$\rho$ & traffic ratio $\lambda_0N/\mu$ & $0.6$\\
\end{tabular}
\end{ruledtabular}
\end{table}

Table~\ref{tab:params} lists the representative parameters. First-passage quantities for one searcher come from the exact arc solutions of Appendix~\ref{app:search}. The gain surface $W(\gamma,p)$ is tabulated on a grid of offsets and reliances in $[0,s_m\sigma]$, and the Gaussian and Poisson averages are taken by quadrature on that table. The two-donor rates $\Lambda_j(\gamma,p_1,p_2)$ are tabulated on a coarser grid of signed offsets and reliances. The reduced dynamics are integrated with adaptive step control. The one-donor individual-based simulation is event driven, and the two-donor simulation advances in small fixed time steps that resolve reversals, service, and departure. Both carry every searcher's position and heading, capture a searcher that crosses a donor with probability $q_0$, and give each fed bee its own scent mark.

The model has two scales, a service time $1/\mu$ and a length $v/\mu$, so Table~\ref{tab:params} maps onto an assay once both are measured. We chose parameters to match the rough scales of arena experiments, in which tens to a hundred or so bees move in a flat arena a few tens of body lengths across and feeding unfolds over tens of minutes~\cite{gharooni-fard_topological_2024}. Bees that walk between a fraction of a body length and a couple of body lengths per second, and are served for several seconds, put the length unit at a few body lengths. The ring is then an arena a few tens of body lengths around, with $\sigma$ and $p_0$ a few body lengths.

\section{First-passage calculations}
\label{app:search}

\subsection{One donor without a cue}
\label{app:single_target}

Place the donor at the origin and let $T_\pm(x)$ be the MFPT from position $x$ with heading $\pm$ at constant reversal rate $\beta$. Integrating the adjoint of Eq.~\eqref{eq:single_stat} in time gives the backward equations
\begin{align}
-1 &= +vT_+' - \beta T_+ + \beta T_- - \kappa\chi_aT_+,\nonumber\\
-1 &= -vT_-' + \beta T_+ - \beta T_- - \kappa\chi_aT_-,
\label{eq:single_backward}
\end{align}
with periodicity and continuity at $x=\pm a$. In the sum $T_s=T_++T_-$ and difference $T_d=T_+-T_-$, the exchange symmetry $T_\pm(x)=T_\mp(-x)$ makes $T_d$ odd and $T_s$ even. Outside the donor, with $y=x-\pi$, the equations give $T_d=-2y/v$ and $T_s=C-2\beta y^2/v^2$. Inside, $T_s''=\alpha^2(T_s-2/\kappa)$ with $\alpha=\sqrt{(2\beta+\kappa)\kappa}/v$, so $T_s=2/\kappa+B\cosh\alpha x$ and $T_d=v\alpha B\sinh(\alpha x)/(2\beta+\kappa)$. Continuity of $T_d$ at $x=a$ fixes $B=2(\pi-a)(2\beta+\kappa)/[v^2\alpha\sinh(\alpha a)]$, continuity of $T_s$ fixes $C$, and averaging $T_s/2$ over the ring gives
\begin{align}
\bar T(\beta)
= \frac{2 \pi - a}{\pi \kappa}
+ \frac{(\pi - a)^2 \alpha}{\pi \kappa \tanh(\alpha a)}
+ \frac{2 \beta (\pi -a)^3}{3 \pi v^2}.
\label{eq:Tbar_beta}
\end{align}
The first term is independent of $\beta$ and the third grows linearly. The second grows as well, since $\mathrm{d}[\alpha/\tanh(\alpha a)]/\mathrm{d}\beta$ has the sign of $h(u)=\tanh u-u\,\mathrm{sech}^2u$ at $u=\alpha a$, and $h(0)=0$ with $h'(u)=2u\,\mathrm{sech}^2u\tanh u>0$. The averaged MFPT is therefore minimized at $\beta=0$.

\subsection{Arc solution with a cue}
\label{app:arcs}

With $\Delta=\beta_--\beta_+$ and $\Sigma=\beta_-+\beta_+$, the backward equations away from the donor read
\begin{align}
vT_d'+\Delta T_d=-2,
\qquad
vT_s'=\Sigma T_d,
\label{eq:backward_sd}
\end{align}
so a response law enters only through $\Delta$ and $\Sigma$. On a flank of the tent, Eq.~\eqref{eq:turning_rate} gives $\beta_\pm=\beta_0(1\mp w)$ with the constant $w=\gamma c_j'/s_m$, $|w|=\gamma/s_m\sigma$, and
\begin{align}
\Delta=2\beta_0 w,
\qquad
\Sigma=2\beta_0,
\label{eq:DS}
\end{align}
so that $\Delta\to2\beta_0$ at the ceiling $\gamma\to s_m\sigma$, where a searcher heading up the gradient never reverses. On an arc of constant $\Delta\neq0$, in a local coordinate $y$,
\begin{align}
T_d(y)&=A\,e^{-\Delta y/v}-\frac{2}{\Delta},\nonumber\\
T_s(y)&=B-\frac{\Sigma}{\Delta}\left(A\,e^{-\Delta y/v}+\frac{2y}{v}\right),
\label{eq:arc}
\end{align}
and $T_d$ is affine and $T_s$ quadratic when $\Delta=0$. The three tent kinks and the donor cut the ring into four arcs with eight constants, fixed by continuity at the kinks and by the two donor conditions
\begin{align}
T_d(0^+)-T_d(0^-)&=\eta\left[T_s(0^+)+T_s(0^-)\right],\nonumber\\
T_s(0^+)-T_s(0^-)&=\eta\left[T_d(0^+)+T_d(0^-)\right],
\label{eq:jump}
\end{align}
with $\eta=\tanh(\overline\kappa/2v)=q_0/(2-q_0)$. At $\gamma=0$ the solution is independent of the offset,
\begin{align}
\bar T(0)=\frac{\pi\left(2-q_0\right)}{vq_0}+\frac{2\beta_0\pi^2}{3v^2},
\label{eq:gamma0}
\end{align}
the ballistic time $1/\lambda_0$ plus a small reversal penalty. The arc solution agrees with event-driven simulations of many walkers to within sampling error at every absorptivity, offset, and reliance tested.

\begin{figure*}[t]
\centering
\includegraphics[width=\textwidth]{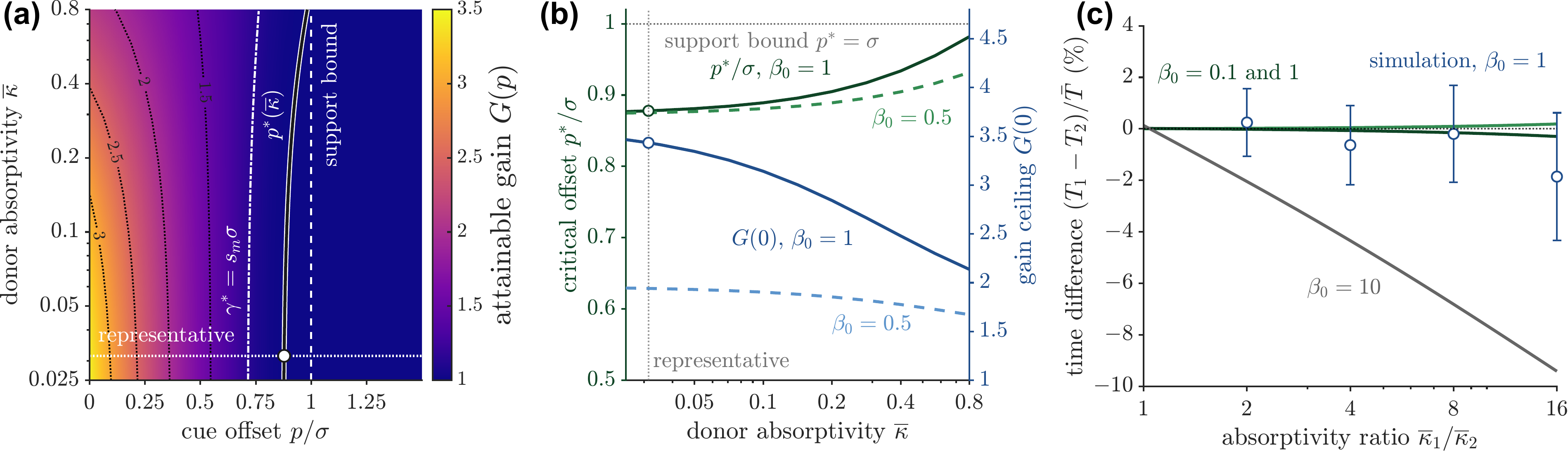}
\caption{\protect\justifying The critical offset is set by geometry, while the gain ceiling depends on how hard the donor is to find. {\bf (a)}~Attainable gain $G(p)$ over cue offset $p/\sigma$ and donor absorptivity $\overline\kappa$ at $\beta_0=1$, with black dotted contours at levels of $G$. The solid curve is the critical offset $p^\ast$, the dash-dotted curve the offset below which the optimal reliance sits on the ceiling $s_m\sigma$, the dashed line the support bound $p=\sigma$, and the white dotted line the absorptivity of Table~\ref{tab:params}. {\bf (b)}~Critical offset $p^\ast/\sigma$ (green, left axis) and gain ceiling $G(0)$ (blue, right axis) against $\overline\kappa$ at $\beta_0=1$ (solid) and $\beta_0=0.5$ (dashed). {\bf (c)}~Difference of the conditional first-passage times to two donors without cues, $(T_1-T_2)/\bar T$, against their absorptivity ratio, with the donors at $\pm\pi/2$ and the geometric mean of their absorptivities fixed at the absorptivity of Table~\ref{tab:params}, from the arc solution at $\beta_0=0.1$, $1$, and $10$ (curves) and simulation at $\beta_0=1$ ($10^5$ walkers, bars $\pm1.96$ standard errors). All panels use $v=1$, and (a) and (b) use $\sigma=0.8$ and $s_m=1$.}
\label{fig:app_search}
\end{figure*}

Figure~\ref{fig:app_search} sweeps donor absorptivity. The critical offset stays below the support bound $\sigma$ and approaches it as absorption strengthens, while the gain ceiling falls. Optimal reliance sits on the ceiling at small offsets and falls to zero at the critical offset $p^\ast$.

\label{app:ceiling}

\textit{Transit from an offset peak.} Under strong reliance a searcher is held near the displaced peak, and discovery requires a transit from the peak at $x=p$ to the donor at $x=0$ across the rising flank, on which the reversal rates are constant. Treating the peak as reflecting and the donor as partially absorbing, so that $T_-(0)=(1-q_0)T_+(0)$, Eq.~\eqref{eq:arc} gives a heading-averaged transit time carrying the factor $e^{\Delta p/v}$, with $\Delta$ and $\Sigma$ from Eq.~\eqref{eq:DS}. As reliance vanishes it reduces to $\beta_0p^2/v^2+p(2-q_0)/(q_0v)$, a diffusive return plus the extra crossings that partial absorption demands. The factor $e^{\Delta p/v}$ is the transit penalty, and since $\Delta\le2\beta_0$ by Eq.~\eqref{eq:DS}, it stays below $e^{2\beta_0p/v}$, so the tradeoff is gentle.

\subsection{Two donors}
\label{app:two_targets}

A donor of half-width $a$ and rate $\overline\kappa_j/2a$ absorbs at a rate of order $a^{-1}$, which dominates reversal across the shrinking region, and integrating across it gives the matching condition
\begin{align}
\pi_1^\pm(x_j^+)-\pi_1^\pm(x_j^-)
=\pm\eta_j\left[\pi_1^\pm(x_j^+)+\pi_1^\pm(x_j^-)-2\delta_{1j}\right].
\label{eq:split_jump}
\end{align}
For donors at $x_1=0$ and $x_2=\pi$ the sum $\pi_1^++\pi_1^-$ is affine and the difference constant on each arc, and with $\mathcal Z=2\pi\beta\eta_1\eta_2+v\eta_1+v\eta_2$,
\begin{align}
\pi_1(x)=\begin{cases}
\eta_1\left[2\beta(\pi-x)\eta_2+v\right]/\mathcal Z, & 0<x<\pi,\\
\eta_1\left[2\beta(x-\pi)\eta_2+v\right]/\mathcal Z, & \pi<x<2\pi,
\end{cases}
\label{eq:pi1}
\end{align}
As $\beta\to0$ the split is $\eta_1/(\eta_1+\eta_2)$ from every start, and as $\beta\to\infty$ it becomes linear in the starting position, with average $1/2$. The arc solution agrees with simulations of many walkers across the crossover. The transform $\hat g_j^\pm(x,s)$ of Eq.~\eqref{eq:conditional_times} satisfies the homogeneous arc equations with donor conditions $\hat g_j^\pm(x_k^\mp)=q_{0,k}\delta_{jk}+(1-q_{0,k})\hat g_j^\pm(x_k^\pm)$. Unequal absorptivities shift the split toward the more discoverable donor, while the conditional discovery times remain nearly equal over a broad range of absorptivity ratios. These times separate appreciably only when reversals are frequent [Fig.~\ref{fig:app_search}(c)].

\label{app:two_cues}

\textit{Two cues.} With a cue at donor 1 alone, recruitment increases $\Lambda_1$ and decreases $\Lambda_2$. Searchers reach the unadvertised donor less often, but those that do arrive sooner, consistent with simulations. With cues at both donors the total rate lies well below the product form $\Lambda^0\sum_jG(p_j)$, with $\Lambda^0$ the no-cue rate per donor, and the deficit is largest when one cue lies beyond $p^\ast$, since a single shared reliance cannot ignore the misaligned cue while following the other. The ratio $\Lambda_1/\Lambda_2$ is reproduced far better than the total when both cues are well aimed. Averaged over scenter configurations at both donors,
\begin{align}
\widetilde\Lambda_j(\bar F_1,\bar F_2)={}&\sum_{k_1,k_2\ge0}\mathcal P(k_1\mid\bar F_1)\,\mathcal P(k_2\mid\bar F_2)\nonumber\\
&\times\left\langle\Lambda_j(p_1,p_2)\right\rangle_{k_1,k_2},
\label{eq:two_averaged}
\end{align}
where $\mathcal P(k\mid\bar F)=e^{-\bar F}\bar F^k/k!$ is the probability that $k$ bees scent at a donor whose mean scenting population is $\bar F$. The inner average $\langle\cdot\rangle_{k_1,k_2}$ runs over cue offsets $p_i\sim\mathcal N(0,p_0^2/k_i)$, with cue $i$ absent when $k_i=0$. The two-donor table covers offsets within the cue support, $|p|<\sigma$, and Gaussian mass beyond it is assigned to the no-cue state. This omits the harm that a misaligned cue does to a searcher following it, which the one-donor gain retains, so the two-donor rates are somewhat optimistic for noisy cues. The reliance maximizes $\widetilde\Lambda_1+\widetilde\Lambda_2$ after the average.

\section{Recruitment gain with a stochastic offset}
\label{app:recruit}

\subsection{Peak offset of an aggregate plume}
\label{app:offset}

Let scenter $i$ stand at $x_j+\xi_i$ and cast a unit tent of half-width $\sigma$, with the $\xi_i$ independent, symmetric about zero, and of scale $s$. With $y=x-x_j$ the aggregate slope is
\begin{align}
c_k'(x)=\frac{1}{\sigma}\Big[n(y,y+\sigma)-n(y-\sigma,y)\Big],
\label{eq:count}
\end{align}
where $n(I)$ counts scenters in the interval $I$, so the cue rises where more scenters lie ahead within one footprint than behind. The peak of a narrow cluster is the sample median, whereas a smooth plume would place it near the sample mean, of variance $s^2/k$. The model takes the mean, $p\sim\mathcal N(0,p_0^2/k)$, and the individual-based simulations place the peak at the mean of the scent marks, so reduction and simulation share one convention.

For these parameters the scatter $s=p_0$ is comparable to $\sigma$, so a literal summed-tent peak follows neither idealization. Its variance grows faster than $1/k$ because a few outlying scenters can build a secondary crest [Fig.~\ref{fig:app_gain}(c)], so for $k\ge2$ it sharpens more slowly than the model assumes. Its steeper flanks cannot be exploited by a searcher whose reliance is capped at $s_m\sigma$. Carrying the measured variance through Eq.~\eqref{eq:averaged_gain} in place of $p_0^2/k$ bounds what the convention costs. The elasticity stays single peaked, with a lower peak at a smaller scenting population. The two conventions agree at $k=1$, so the value of the first scenter, the slope at the origin in Eq.~\eqref{eq:g1}, and the absence of a minimum scenter number do not depend on the convention, while the benefit of a large scenting population is an upper bound.

\subsection{The averaged recruitment gain}
\label{app:gain}

Hungry bees are fed at rate $\nu=\mu U/(U+F)$ and fed bees depart at rate $\delta$, which is the $M/M/\infty$ occupancy process with a Poisson stationary law of mean $\nu/\delta$~\cite{kleinrock_queueing_1975}. Because $\nu$ depends on the state, the law is quasi-stationary. We average rates rather than times because discovery at a fixed cue is nearly Poisson (Appendix~\ref{app:gap}) and the cue turns over far faster than the search time $1/\lambda_0$ at the optimum, so accumulated capture self-averages, whereas averaging times would understate the benefit. A searcher fixes $\gamma$ once and cannot condition it on a configuration redrawn many times within its search, so the gain is $\max_\gamma\langle W\rangle$, not $\langle\max_\gamma W\rangle$.

\begin{figure*}[t]
\centering
\includegraphics[width=\textwidth]{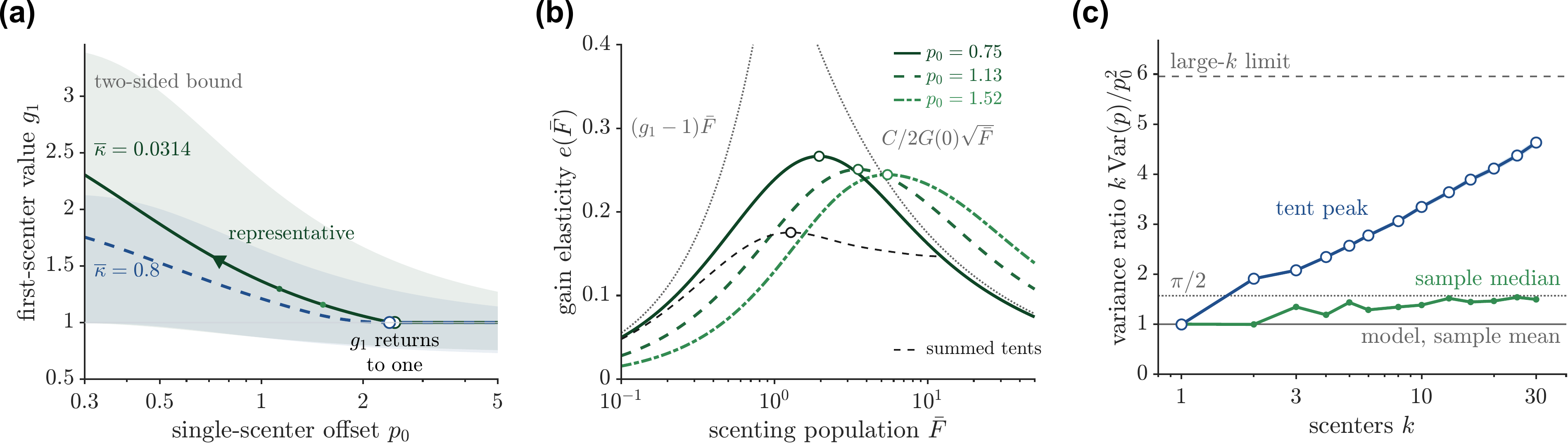}
\caption{\protect\justifying Recruitment grows from the first scenter and approaches its ceiling slowly, while a literal sum of tents sharpens more slowly than the model assumes. {\bf (a)}~Value of the first scenter $g_1$ against $p_0$ at $\overline\kappa=0.0314$ (solid) and at $\overline\kappa=0.8$ (dashed), with the two-sided bound on $g_1$ (shading). Open circles mark where $g_1$ returns to one. {\bf (b)}~Gain elasticity $e(\bar F)$ for $p_0=0.75$, $1.13$, $1.52$, peaks marked ($e=0.267$ at $\bar F=1.96$ for $p_0=0.75$), with the limits Eq.~\eqref{eq:elasticity_limits} at $p_0=0.75$ (dotted). The black dashed curve is $p_0=0.75$ with the measured peak variance of a literal sum of tents in place of $p_0^2/k$, shown to $\bar F=12$, where the measured counts cover the Poisson weight. {\bf (c)}~Scaled variance $k\,\mathrm{Var}(p)/p_0^2$ of the peak of the summed-tent cue from $k$ scenters with Gaussian positions (blue, band $\pm1.96$ standard errors), against the sample median (green), the model value $1$ of the sample mean, and the large-$k$ limit of the tent peak.}
\label{fig:app_gain}
\end{figure*}

With the Poisson weights $\mathcal P_k=\mathcal P(k\mid\bar F)$ of Eq.~\eqref{eq:two_averaged}, which satisfy $\mathcal P_k'=\mathcal P_{k-1}-\mathcal P_k$,
\begin{align}
\frac{\pd}{\pd\bar F}\sum_k\mathcal P_k\,a_k(\gamma)=\sum_{k\geq0}\mathcal P_k\left[a_{k+1}(\gamma)-a_k(\gamma)\right],
\label{eq:increment}
\end{align}
the expected value of one more scenter at a Poisson incumbent count. The computed surface is even and nonincreasing in $|p|$ at each $\gamma$, so $a_k(\gamma)$ is nondecreasing in $k$ for $k\ge1$, and since $\gamma=0$ gives $a_k=1$ for all $k$, $\widetilde G\ge1$. The maximizing reliance sits at the ceiling $s_m\sigma$ for every $\bar F>0$ and every offset we use, so $\widetilde G$ is nondecreasing. At $\bar F=0$ every reliance gives the same value, so the right derivative of the maximum is the largest partial derivative over all reliances, and Eq.~\eqref{eq:increment} gives Eq.~\eqref{eq:g1}, while for $\bar F>0$ the maximizer is unique and the envelope theorem gives $\widetilde G'$ without differentiating $\gamma^\ast$.

Whether the first scenter pays is set by a competition between the benefit of a well-placed cue and the harm of a misplaced one. With $W\ge W_{\min}$ everywhere, $W\le G(0)$, and $\Phi_\ast=2\Phi(p^\ast/p_0)-1$ the probability that a lone scenter lands within $p^\ast$ of the donor, where $\Phi$ is the standard normal distribution function, $g_1$ lies between $1-(1-W_{\min})(1-\Phi_\ast)$ and $1+(G(0)-1)\Phi_\ast$, the upper bound because $W\le1$ beyond $p^\ast$ at every reliance, and the lower because $W\ge1$ within $p^\ast$ at small reliance. The harm saturates at $1-W_{\min}$ while the benefit vanishes with $\Phi_\ast$, so beyond a finite offset spread, $g_1=1$ and recruitment waits for a second scenter. That spread is a few cue half-widths at both absorptivities [Fig.~\ref{fig:app_gain}(a)], and the three offsets of Fig.~\ref{fig:lingering} all lie below it, so each has $g_1>1$.

At large $\bar F$ the offset distribution concentrates. At the ceiling reliance the gain surface has a cusp at zero offset, $W\simeq G(0)-c_1|p|$, because a peak displaced to either side moves the donor onto a flank whose bias points away from it. Averaging $|p|$ over $\mathcal N(0,p_0^2/k)$ and $k^{-1/2}$ over the Poisson law gives $\widetilde G(\bar F)=G(0)-C/\sqrt{\bar F}$ with $C=c_1p_0\sqrt{2/\pi}$, and the elasticity behaves as
\begin{align}
e(\bar F)&\sim\left(g_1-1\right)\bar F, &&\bar F\to0,\nonumber\\
e(\bar F)&\sim\frac{C}{2G(0)\sqrt{\bar F}}, &&\bar F\to\infty,
\label{eq:elasticity_limits}
\end{align}
so it is single peaked, with a peak at intermediate $\bar F$ [Fig.~\ref{fig:app_gain}(b)].

\section{Abandonment limits, validity, and robustness}
\label{app:validity}

\subsection{The two abandonment limits}
\label{app:limits}

\begin{figure*}[t]
\centering
\includegraphics[width=\textwidth]{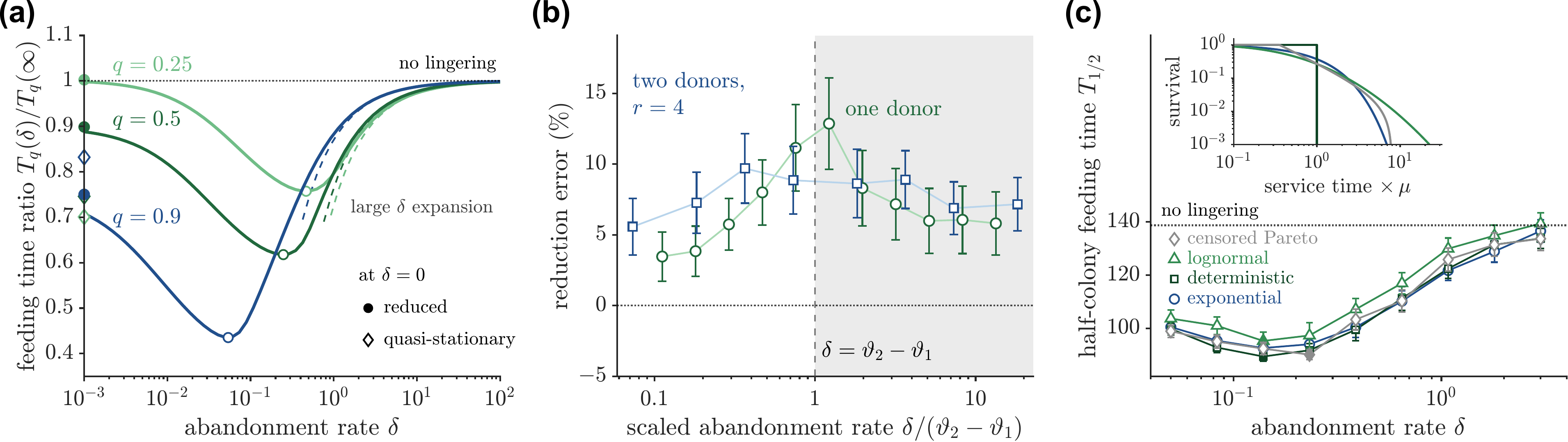}
\caption{\protect\justifying The optimum is interior, the reduction fails where the cue turns over faster than searchers relax, and the service law barely matters. {\bf (a)}~Feeding time relative to immediate departure from the reduced dynamics at $q=0.25$, $0.5$, $0.9$, with the large-$\delta$ expansion Eq.~\eqref{eq:large_delta} (dashed) and optima (open circles). Markers on the left axis are $T_q(0)$ from the reduced dynamics (filled) and from the quasi-stationary closure (open). {\bf (b)}~Relative error of the reduction against the individual-based model with $\delta$ scaled by the spectral gap, one donor ($150$ realizations) and two donors at $r=4$ ($200$ realizations), bars $\pm1.96$ standard errors. Shading marks abandonment rates above the spectral gap. {\bf (c)}~Half-colony feeding time under four service laws of equal mean ($80$ realizations per point), exponential, deterministic, lognormal (coefficient of variation $1.99$), and censored Pareto (exponent $1.21$, coefficient of variation $1.05$), with filled markers at each optimum. The inset shows the survival functions of the four laws. Parameters as in Fig.~\ref{fig:lingering}.}
\label{fig:app_limits}
\end{figure*}

Quasi-stationarity gives $F=A/\delta$ and $U=AF/(\mu-A)$ with $A=\lambda_0\widetilde G(F)S$, and conservation turns the stopping condition into $S+U\le(1-q)N$, so
\begin{align}
T_q=\int_{S_q}^{N}\frac{\mathrm{d}S}{\lambda_0\widetilde G(F(S))\,S},
\qquad S_q+U(S_q)=(1-q)N.
\end{align}
As $\delta\to\infty$ the integrand is $1/\lambda_0S$ and $S_q=(1-q)N$, so the integral is $\ell_q/\lambda_0$. At large but finite $\delta$, $F\simeq\lambda_0S/\delta$ and $1/(\lambda_0\widetilde GS)\simeq(\lambda_0S)^{-1}[1-(g_1-1)\lambda_0S/\delta]$ with $\widetilde G=1+(g_1-1)F$, so the integrand contributes $-(g_1-1)qN/\delta$. The congestion $U\simeq A_qF_q/(\mu-A_q)$, with $F_q=A_q/\delta$, moves the endpoint to $S_q\simeq(1-q)N-U$ and adds $U/A_q=F_q/(\mu-A_q)$. Together these give Eq.~\eqref{eq:large_delta}. Extrapolating the reduced dynamics in $1/\delta$ reproduces the bracket at every fed fraction [Fig.~\ref{fig:app_limits}(a)]. At two donors the benefit term is unchanged and the cost becomes a sum over donors.

At $\delta=0$ no bee departs and the fed count rises monotonically, so using $F$ as the clock, $\mathrm{d}t=\mathrm{d}F\,(U+F)/\mu U$, we obtain
\begin{align}
T_q(0)=\frac{qN}{\mu}+\frac{1}{\mu}\int_0^{qN}\frac{F}{U(F)}\,\mathrm{d}F,
\label{eq:T0}
\end{align}
exact along any trajectory of Eq.~\eqref{eq:mean_field}. The feeding time is the pure service time inflated by the dilution of the service lottery. The quasi-stationary closure cannot evaluate this integral, because without departures the fed population never relaxes and $\dot U$ remains comparable to the arrival rate. Retaining $\dot U$ at small traffic ratio, with $\mu U/F\simeq\dot F$ and $\dot F=A-\dot U$, gives without recruitment
\begin{align}
\frac{T_q(0)}{T_q(\infty)}=1+\rho\,\frac{q}{\ell_q}+O(\rho^2)>1.
\label{eq:T0_ratio}
\end{align}
Integrating the reduced dynamics confirms the slope $q/\ell_q$ at small $\rho$ and gives a ratio above one at every traffic ratio and fed fraction we sample, with a substantial penalty at these parameters. With recruitment, the endpoint ratio falls but stays well above the optimum at every fed fraction we sample [Fig.~\ref{fig:app_limits}(a)], so the minimum is interior.

\subsection{Search closure and the coherence bound}
\label{app:gap}

For $N$ independent searchers with first-passage density $f(t)$ and survival probability $Q(t)=\int_t^\infty f\,\mathrm{d}t'$, $\dot S/S=-f/Q$, and the reduction $\dot S=-\lambda S$ is exact only for an exponential $f$. The product of the long-time decay rate $\lambda_1$ and $\bar T$ tests the tail, which governs the end of a run, and the coefficient of variation tests the spread. For a ballistic searcher both are essentially unity at our weak donor, although arrivals occur only at discrete crossing times spaced by a circuit. Reversal randomizes those crossings and spreads arrivals continuously, and at our reversal rate both lie close to unity over the offsets and reliances we use, with a gain defined from $\lambda_1$ agreeing closely with $W$. The width of Eq.~\eqref{eq:coherence_window} is $(2-q_0)/2q_0$, which is large for these parameters. The relaxation rate onto a cue is the spectral gap of the biased search generator. With $\phi_s$ and $\phi_d$ the sum and difference of the heading-resolved eigenfunctions, the eigenproblem on each arc is
\begin{align}
v\phi_d'=-\Delta\phi_d-\vartheta\phi_s,
\qquad
v\phi_s'=\left(\Sigma-\vartheta\right)\phi_d,
\end{align}
with the donor conditions of Eq.~\eqref{eq:jump}, and $\vartheta$ is an eigenvalue when the monodromy matrix around the ring has eigenvalue one. At zero reliance the reciprocal of the smallest root, $1/\vartheta_1$, agrees with $\bar T(0)$ from Eq.~\eqref{eq:gamma0}. At full reliance the gap $\vartheta_2-\vartheta_1$ is smaller for two plumes than for one. At one donor the relative error of the reduction is modest for $\delta$ below the gap, peaks just past it, and falls again at large $\delta$ [Fig.~\ref{fig:app_limits}(b)]. At two donors it stays roughly constant across the sweep, so the two-donor mismatch is not a coherence effect alone.

\subsection{Non-exponential service}
\label{app:service}

Measured trophallactic durations are broad and heavy tailed~\cite{choi_individual_2020,gernat_automated_2018}. We simulated the individual-based model with service intervals of equal mean $1/\mu$ drawn from exponential, deterministic, lognormal, and Pareto (censored at the longest measured duration) laws. The service clock idles while the aggregation is empty, and a fed bee drawn for service consumes a full interval while feeding no one. On the coarser grid of this sweep, every law yields an interior optimum within one grid step of the exponential one, with similar speedups [Fig.~\ref{fig:app_limits}(c)].

\bibliographystyle{apsrev4-2}
\bibliography{refs}

\end{document}